\documentclass[a4paper,11pt]{article}
\pdfoutput=1
\usepackage{jcappub}
\usepackage{epsf}
\usepackage[utf8]{inputenc}
\usepackage{amstext}
\usepackage{graphicx}
\usepackage{amssymb}
\usepackage{amsmath}
\usepackage{multirow}
\usepackage{amsthm}

\title{\boldmath $H_0$ from cosmic chronometers and Type Ia supernovae, with Gaussian Processes and the novel Weighted Polynomial Regression method}

\author[a]{Adri\`a G\'omez-Valent}
\author[b]{and Luca Amendola}

\affiliation[a]{Departament de F\'isica Qu\`antica i Astrof\'isica, and Institut
de Ciències del Cosmos, Univ.\ de Barcelona, Av.\ Diagonal 647, E-08028 Barcelona,
Catalonia, Spain}
\affiliation[b]{Institut für Theoretische Physik, Ruprecht-Karls-Universit\"{a}t Heidelberg, Philosophenweg 16, 69120 Heidelberg, Germany}

\emailAdd{adriagova@fqa.ub.edu}
\emailAdd{l.amendola@thphys.uni-heidelberg.de}

\abstract{%
In this paper we present new constraints on the Hubble parameter $H_0$ using: (i) the available data on $H(z)$ obtained from cosmic chronometers (CCH); (ii) the Hubble rate data points extracted from the supernovae of Type Ia (SnIa) of the Pantheon compilation and the Hubble Space Telescope (HST) CANDELS and CLASH Multy-Cycle Treasury (MCT) programs; and (iii) the local HST measurement of $H_0$ provided by Riess {\it et al.} (2018), $H_0^{{\rm HST}}=(73.45\pm1.66)$ km/s/Mpc. Various determinations of $H_0$ using the Gaussian processes (GPs) method and the most updated list of CCH data have been recently provided by Yu, Ratra \& Wang (2018). Using the Gaussian kernel they find $H_0=(67.42\pm 4.75)$ km/s/Mpc. Here we extend their analysis to also include the most released and complete set of SnIa data, which allows us to reduce the uncertainty by a factor $\sim 3$ with respect to the result found by only considering the CCH information. We obtain $H_0=(67.06\pm 1.68)$ km/s/Mpc, which favors again the lower range of values for $H_0$ and is in tension with $H_0^{{\rm HST}}$. The tension reaches the $2.71\sigma$ level. We round off the GPs determination too by taking also into account the error propagation of the kernel hyperparameters when the CCH with and without $H_0^{{\rm HST}}$ are used in the analysis. In addition, we present a novel method to reconstruct functions from data, which consists in a weighted sum of polynomial regressions (WPR). We apply it from a cosmographic perspective to reconstruct $H(z)$ and estimate $H_0$ from CCH and SnIa measurements. The result obtained with this method, $H_0=(68.90\pm 1.96)$ km/s/Mpc, is fully compatible with the GPs ones. Finally, a more conservative GPs+WPR value is also provided, $H_0=(68.45\pm 2.00)$ km/s/Mpc, which is still almost $2\sigma$ away from $H_0^{{\rm HST}}$.}

\arxivnumber{arXiv:1802.01505}

\begin{document}
\maketitle
\flushbottom

\section{Introduction}
The Hubble constant, $H_0$, is a pivotal quantity in cosmology, since it is used to construct the most basic time and distance cosmological scales. It was measured by the very first time by E. Hubble in 1929, who estimated it to be around $500$ km/s/Mpc \cite{Hubble1929}. Of course we know now that this first determination was exceedingly large, namely a factor $\sim 7$ above the true value. Present measurements agree that it must fall in the approximate range 65-75 km/s/Mpc, but the situation regarding the available observational values of $H_0$ is disquieting. The reason is that there currently exists a significant tension between different measurements and determinations of this parameter. For instance, there is a big discrepancy between values obtained from local measurements depending on how the distances to the supernovae of Type Ia (SnIa) used in the analysis are calibrated. As an example we can compare the values inferred from the local Hubble Space Telescope (HST) measurement of \cite{RiessH02018}, $H_0^{\rm HST}=(73.45\pm 1.66)$ km/s/Mpc, with the one provided in \cite{TammannReindl2013}, $H_0=(63.7\pm 2.3)$ km/s/Mpc. The tension between these values is not small, around $3.4\sigma$. The latter prefers the lower range and is more aligned with the investigations carried out by Allan Sandage and collaborators for many years \cite{OnSandage}, whereas the former lies in the higher range, follows the line of the previous analyses \cite{Riess2011,RiessH02016,RiessH02017} and is also supported by other works, see e.g. \cite{Freedman2012,Zhang2017,Feeney2017,Cardona2017,Casertano2017,SuhailDhawan2018}. Recent strong lensing \cite{Bonvin2017} and HII galaxy \cite{WangMeng2017} measurements also favor the higher range of $H_0$-values. It is appropriate to mention too the recent determination $H_0=(71.17\pm 2.50)$ km/s/Mpc \cite{JangLee2017}, obtained (as the one from \cite{TammannReindl2013}) by using the tip of the red giant branch distances. It loosens the tension with the HST value. On the contrary, the authors of the dedicated study \cite{Lin2017} find some indication that the local HST measurement of $H_0$ is an outlier upon performing a consistency test considering the constraints on the Hubble parameter that are obtained from a large variety of sources (including the one from Ref. \cite{Bonvin2017}), thus favoring a systematics-based explanation of the $H_0$-tension. In addition, we must also comment on the well-known tension between the HST value from \cite{RiessH02018} and the one from Planck's collaboration \cite{Planck2016}, $H_0^{\rm P16}=(66.93\pm 0.62)$ km/s/Mpc, derived from the analysis of the TT,TE,EE+SIMlow cosmic microwave background (CMB) data by assuming the $\Lambda$CDM cosmological model as the true one. In this case there is a $3.7\sigma$ tension between these two determinations. A $\sim 3\sigma$ tension is also found when comparing the HST value with the Planck one provided in the preceding paper \cite{PlanckAde}. At this moment the mismatch between Planck's values and some of the local measurements of $H_0$ cannot be explained by any crucial systematic error in the data. If it does exist, it has not been detected yet, see Refs. \cite{Aylor2017,Addison2017,Follin2017}. It cannot be solved by the ``Hubble bubble'' hypothesis neither. The latter states that we live in a locally underdense region of the universe which makes us measure a higher value of $H_0$ than compared to the global one. Although this idea is interesting, it has proven to be highly improbable \cite{Marra2013,Wojtak2014}. Certainly, it is quite shocking that in this age of precision cosmology, twenty years after the discovery of the speeding-up of the universe \cite{SNIaRiess,SNIaPerl} and almost ninety years after the discovery of its expansion \cite{Hubble1929}, we still lack a resonant determination of $H_0$, the first ever measured cosmological parameter.     

Some attempts at loosening this tension from a theoretical perspective have been made by several authors, assuming new physics scenarios. For instance, by exploring extensions of the vanilla $\Lambda$CDM parameter space \cite{Melchiorri2016,Melchiorri2017}, by modifying the early-time physics adding some kind of dark radiation \cite{VerdeRiessH0} or early dark energy (DE) \cite{MortsellDhawan2018}, by allowing DE interactions with dark matter (DM) \cite{Melchiorri2017b}, or with the aid of massive neutrinos \cite{Wyman2014}. See also the recent Ref. \cite{Poulin2018}. It is worth mentioning that when the large scale structure formation data is also considered in the fitting analysis of the concordance model the predicted value of $H_0$ is dragged away from Planck's best-fit estimate too, and enters in conflict with both, the Planck and the HST values \cite{PLB2017}. In \cite{PLB2017,EPL2017,MNRAS2018} it is found that some dynamical DE models can loosen the $\sigma_8$-tension in a very effective way by respecting at the same time the $H_0$-value preferred by Planck, so in these scenarios the lower range for the Hubble parameter is clearly favored. Previously it was shown that they provide an overall fit to the current cosmological data significantly better than the $\Lambda$CDM \cite{ApJL2015,ApJ2017,PRD2017,GomezValentThesis,EPL2018x3}.

In view of the crudeness of the $H_0$-tension exposed above, it is very important to obtain alternative estimates as reliable as possible for this parameter. Median statistics have been largely applied to get some constraints by using also historical compilations of $H_0$-measurements \cite{Gott2000,ChenRatra2011,Bethapudi2017}, as Huchra's compilation\footnote{\href{https://www.cfa.harvard.edu/~dfabricant/huchra/hubble.plot.dat}{https://www.cfa.harvard.edu/~dfabricant/huchra/hubble.plot.dat}.} and updated extensions of it. Although this method somehow minimizes the effect of outliers, which is of course very welcome and something that the usual weighted mean does not do, it does not incorporate the information of the individual uncertainties of the data set elements, and therefore loses part of the information. Despite this issue, it proves to be a very robust method. The central values for $H_0$ obtained by using it tend to lie in the low range, but their uncertainties are still quite large and this fact makes them compatible with the local determinations from \cite{RiessH02018,Riess2011,RiessH02016,RiessH02017}. For instance, the authors of \cite{ChenRatra2011} obtain $H_0=(68.0 \pm 5.6)$ km/s/Mpc at $95\%$ c.l..

The birth of gravitational-wave multi-messenger astronomy has also come hand in hand with new measurements of the Hubble parameter. The detection of the gravitational wave and the electromagnetic counterpart produced by the merger of the binary neutron-star system GW170817 \cite{GWevent} has allowed to use this event as a standard siren and measure $H_0$ \cite{GWH0,GWH0second}. The value reported in \cite{GWH0} is $H_0=70.0^{+12.0}_{-8.0}\,{\rm km/s/Mpc}$, which lies between the Planck and the HST values. The one presented in \cite{GWH0second} is higher, $H_0=75.5^{+11.6}_{-9.6}\,{\rm km/s/Mpc}$. Although these constraints are still quite poor and are unable to discriminate between the two values in dispute, it is worth to remark that they have been obtained without using any form of cosmic distance ladder, and therefore are free of systematic errors that could affect the more standard astronomical determinations of cosmological distances. Nevertheless, binary neutron-star standard sirens can play an important role in the next decade. In Ref. \cite{Feeney2018} the authors demonstrate that a typical sample of $\sim 50$ of these objects can independently arbitrate the $H_0$-tension, and this will be possible in the following years thanks to the LIGO and Virgo experiments. Other measurements lying in between the HST and Planck's values are e.g. \cite{Efstathiou2014,BasilakosH0,Cantiello2018}. They are actually compatible with both, Planck and the HST, at $1\sigma$.

In this work we extend the Bayesian ``non-parametric'' analysis carried out in the recent paper \cite{YuRatraWang2017} by including the effect of the most complete and updated data on SnIa provided in \cite{Riess2017}. We will see how we can convert in a consistent way the Hubble rates that summarize the information of these SnIa into data on the Hubble function $H(z_i)$ at the effective redshifts $z_i$ considered in \cite{Riess2017}, and how the low uncertainties of the SnIa data allow us to reduce the error of our determination of $H_0$ by a factor $\sim 3$ with respect to the case with no SnIa data, when only cosmic chronometers (CCH) are considered. We find values that lie mostly in the lower range of $H_0$, supporting the preferred band by Planck \cite{Planck2016,PlanckAde}, and also other recent studies as e.g. \cite{Lin2017,PLB2017,YuRatraWang2017,Feeney2018}. We also reanalyze the results obtained with the Gaussian Processes (GPs) method for the CCH and CCH+$H_0^{{\rm HST}}$ data sets, by properly accounting for the error propagation of the hyperparameters entering the kernel that controls the correlations between the points of the reconstructed function $H(z)$. We will show that this fact results in a slight lowering of the central values of $H_0$ and an increase of the standard deviations with respect to the values inferred without propagating the hyperparameters' errors. Despite this, the corrected values are still compatible at $1\sigma$ with the uncorrected ones, so no important differences are found. We complement our analysis in the context of the GPs with a detailed study on the impact that has the choice of the stellar population synthesis model used to extract the CCH $H(z_i)$-data on our results.

We also present a novel method based on polynomial regression and Occam's razor principle to reconstruct continuous functions from a list of data points. We apply it to the study of $H(z)$ and obtain new determinations of $H_0$ by considering various data set configurations. 

The layout of this paper is as follows. In Sect. \ref{sect:DataSets} we briefly describe the data sets that will be used in the current analysis. In Sect. \ref{sect:GPMethod} we make a quick review of the GPs method and present the basic formulas. We apply this method to the CCH and CCH+Pantheon+MCT data sets, and also study the impact of the local $H_0$-determination \cite{RiessH02018} on our results. Later on we apply again the GPs method to the CCH data in a fully consistent way, in order to obtain the corrected GP determination of $H_0$ by accounting for the error propagation of the kernel hyperparameters. We end Sect. \ref{sect:GPMethod} by studying the impact on our results of potential systematic errors affecting the CCH data, and also analyze the effect of other changes in our CCH data set. In Sect. \ref{sect:WPR} we describe in detail the novel weighted polynomial regression (WPR) method, and apply it to different data sets and under two different cosmographical perspectives to reconstruct $H(z)$ and derive $H_0$. In Sect. \ref{sect:DiscussionConclusions} we finally discuss the results and present our conclusions.


\section{The data sets}\label{sect:DataSets}

We devote this section to present and describe the data that will be used later on in the subsequent analyses based on Gaussian processes (in Sect. 3) and the weighted polynomial regression approach (in Sect. 4). We also provide the corresponding references, together with the model dependencies and assumptions behind these data. This is of course basic to ensure the correct understanding of the scope of our calculations and the range of validity of our results.

\begin{table}[!t]
\begin{center}
\begin{tabular}{| c | c | c |}
\multicolumn{1}{c}{$z$} &  \multicolumn{1}{c}{$H(z)$} & \multicolumn{1}{c}{{\small References}}
\\\hline
$0.07$ & $69.0\pm 19.6$ & \cite{Zhang}
\\\hline
$0.09$ & $69.0\pm 12.0$ & \cite{Jimenez}
\\\hline
$0.12$ & $68.6\pm 26.2$ & \cite{Zhang}
\\\hline
$0.17$ & $83.0\pm 8.0$ & \cite{Simon}
\\\hline
$0.1791$ & $75.0\pm 4.0$ & \cite{Moresco2012}
\\\hline
$0.1993$ & $75.0\pm 5.0$ & \cite{Moresco2012}
\\\hline
$0.2$ & $72.9\pm 29.6$ & \cite{Zhang}
\\\hline
$0.27$ & $77.0\pm 14.0$ & \cite{Simon}
\\\hline
$0.28$ & $88.8\pm 36.6$ & \cite{Zhang}
\\\hline
$0.3519$ & $83.0\pm 14.0$ & \cite{Moresco2012}
\\\hline
$0.3802$ & $83.0\pm 13.5$ & \cite{Moresco2016}
\\\hline
$0.4$ & $95.0\pm 17.0$ & \cite{Simon}
\\\hline
$0.4004$ & $77.0\pm 10.2$ & \cite{Moresco2016}
\\\hline
$0.4247$ & $87.1\pm 11.2$ & \cite{Moresco2016}
\\\hline
$0.4497$ & $92.8\pm 12.9$ & \cite{Moresco2016}
\\\hline
$0.47$ & $89.0\pm 49.6$ & \cite{Ratsimbazafy2017}
\\\hline
$0.4783$ & $80.9\pm 9.0$ & \cite{Moresco2016}
\\\hline
$0.48$ & $97.0\pm 62.0$ & \cite{Stern}
\\\hline
$0.5929$ & $104.0\pm 13.0$ & \cite{Moresco2012}
\\\hline
$0.6797$ & $92.0\pm 8.0$ & \cite{Moresco2012}
\\\hline
$0.7812$ & $105.0\pm 12.0$ & \cite{Moresco2012}
\\\hline
$0.8754$ & $125.0\pm 17.0$ & \cite{Moresco2012}
\\\hline
$0.88$ & $90.0\pm 40.0$ & \cite{Stern}
\\\hline
$0.9$ & $117.0\pm 23.0$ & \cite{Simon}
\\\hline
$1.037$ & $154.0\pm 20.0$ & \cite{Moresco2012}
\\\hline
$1.3$ & $168.0\pm 17.0$ & \cite{Simon}
\\\hline
$1.363$ & $160.0\pm 33.6$ & \cite{Moresco2015}
\\\hline
$1.43$ & $177.0\pm 18.0$ & \cite{Simon}
\\\hline
$1.53$ & $140.0\pm 14.0$ & \cite{Simon}
\\\hline
$1.75$ & $202.0\pm 40.0$ & \cite{Simon}
\\\hline
$1.965$ & $186.5\pm 50.4$ & \cite{Moresco2015}
\\\hline
\end{tabular}
\caption{31 published values of $H(z)$ in [km/s/Mpc] obtained using the differential-age technique, see the quoted references and Sect 2.1. They coincide with the data points on cosmic chronometers used in the analysis of \cite{YuRatraWang2017}. We explicitly show them for convenience.}
\end{center}
\end{table}
%

\subsection{Cosmic chronometers}

\begin{table}[!t]
\centering
\label{PantheonMCTData}
\begin{tabular}{|ccccccc|}
\hline
\multicolumn{1}{|c}{$z$} & \multicolumn{1}{c}{$E(z)$} & \multicolumn{5}{c|}{Correlation matrix} \\ 
\hline
 $0.07$& $0.997\pm0.023$ & $1.00$  &   &   &   &     \\
 $0.20$& $1.111\pm0.020$&  $0.39$ &  $1.00$ &   &   &    \\
 $0.35$& $1.128\pm0.037$ & $0.53$  & $-0.14$  & $1.00$  &   &    \\
 $0.55$& $1.364\pm0.063$ &  $0.37$ &  $0.37$ & $-0.16$  & $1.00$  &   \\
 $0.90$& $1.52\pm0.12$&  $0.01$ &  $-0.08$ &  $0.17$ &  $-0.39$ & $1.00$   \\\hline
\end{tabular}
\caption{Values of the Hubble rate taken from Ref. \cite{Riess2017}. Note by comparing this table with table 6 of this reference that we omit the use of the data point at $z=1.5$. This is because, contrary to the others, the latter does not follow a Gaussian distribution, and therefore it is better not to include it in the GPs analyses, which assume the normality of the input data. Nevertheless, the loss of information is not dramatic at all, the reason being that this data point has a large uncertainty (around $\sim 25\%$ in relative terms), and is the most remote from $z=0$, so the impact on our determination of $H_0$ would be in any case quite mild. This is explicitly checked in Appendix A in the context of some model-dependent analyses.}
\end{table}
%

Spectroscopic dating techniques of passively–evolving galaxies, i.e. galaxies with old stellar populations and low star formation rates, have become a good tool to obtain observational values of the Hubble function at redshifts $z\lesssim  2$ (see \cite{JimenezLoeb2002} and also the references in Table 1). These measurements are independent of the Cepheid distance scale and do not rely on any particular cosmological model, although of course are subject to other sources of systematic uncertainties, as the ones associated to the modeling of stellar ages, which is carried out through the so-called stellar population synthesis techniques (SPS), see Sect. 3.4 for further details. Given a pair of ensembles of passively-evolving galaxies at two different redshifts it is possible to infer $dz/dt$ from observations and under the assumption of a concrete SPS model. Therefore it is possible to compute $H(z)=-(1+z)^{-1}dz/dt$ too, which is the quantity we are interested in. Thus, cosmic chronometers allow us to obtain direct information about the Hubble function at different redshifts, contrary to other probes which do not directly measure $H(z)$, but integrated quantities as e.g. luminosity distances. In Table 1 we list the CCH data points on $H(z_i)$ used in our main analyses, together with their corresponding uncertainties $\sigma_i$. We point out that we have used a diagonal covariance matrix for these data, i.e. $D_{ij}=\sigma_i^2\delta_{ij}$.

In the current analysis we do not make use of data on $H(z)$ extracted from the measurement of baryon acoustic oscillations (BAO) in order to avoid dealing with their cosmological model dependence. In this work we will try to reconstruct the continuous curve of $H(z)$ using statistical methods as free as possible of assumptions on a particular physical cosmological model, and determine $H_0$ from the corresponding reconstructed curves. In this sense, we will try to depart as much as possible from the approach adopted in other works, as e.g. in \cite{CunhaMarassiLima2007,LimaCunha2014,Holanda2014,Lukovic2016,ChenKumarRatra2017,WangXuZhao2017,ZhangHuangLi2018,MiaoHuang2018}, in which $H_0$ is determined in the framework of concrete cosmological models \footnote{We only dedicate Appendix A to a model-dependent study. More concretely, to the fitting analysis of the $\Lambda$CDM and XCDM models with the data sets described in Sect. 2.1 and 2.2.}. The first model assumption that we will take for granted is the homogeneity and isotropy of the universe at large scales. These features are exceedingly sustained by radiation backgrounds as CMB observations, and counts of sources observed at wavelengths ranging from radio to gamma rays. Notice that the measurements of CCH listed in Table 1 have been obtained from galaxies located at different angles in the sky. Under the coverage of the cosmological principle (CP) the dependence of the CCH data on the angle and location of the measured galaxies is removed, and therefore we are legitimated to reconstruct $H(z)$ using all the values of Table 1.
   

\subsection{Pantheon+MCT SnIa compilation}

In this work we use the Hubble rate data points, i.e. $E(z_i)=H(z_i)/H_0$, provided in \cite{Riess2017} for six different redshifts in the range $z\in[0.07,1.5]$ (we omit the one at $z=1.5$ because it is not Gaussian-distributed, see the caption of Table 2). They effectively compress the information of the $\sim 1050$ SnIa at $z<1.5$ that take part of the Pantheon compilation \cite{Pantheon} (which includes the 740 SnIa of the joint light-curve analysis sample \cite{BetouleJLA}), and the 15 SnIa at $z>1$ of the CANDELS and CLASH Multy-Cycle Treasury (MCT) programs obtained by the HST, 9 of which are at $1.5<z<2.3$. The authors of \cite{Riess2017} converted the raw SnIa measurements into $E(z)$ by parametrizing $E^{-1}(z)$ at those six redshifts $z_i$. The integral over $E^{-1}$ that defines the luminosity distance is then obtained by interpolating between $z_i$ with cubic Hermite polynomials. Finally, the overall constant $H_0$ is marginalized away along with the absolute supernovae magnitude, see \cite{Riess2017} for further details. The corresponding values of $E^{-1}(z_i)$ are Gaussian in very good approximation and are shown in Table 6 of \cite{Riess2017}, together with the corresponding correlation matrix. We present their inverse, $E(z_i)$, and the correlation matrix in Table 2, just for completeness. They are also almost perfectly Gaussian. It is important to remark that these values have been obtained by assuming a flat universe (and, again, the CP), and thus are model-dependent in this sense. We know, though, that these are in fact quite reasonable assumptions if our main aim is to use these data points to reconstruct $H(z)$ around the current time. Note e.g. that the TT+lowP+lensing+BAO analysis carried out by the Planck collaboration in \cite{PlanckAde} predicts a current spatial curvature density $\Omega_k^{(0)}=0.000\pm0.005$ at $2\sigma$ c.l., which is fully compatible with the flat universe scenario; or the analysis from \cite{RatraOmegak}, which in this case favors a closed universe, although the central value for $\Omega_k^{(0)}$ is still low, around -0.006 when the model is confronted to the TT,TE,EE+lowP+lensing+BAO data. One can easily check that the relative change on $H(z)$ caused but these tiny deviations from flatness is really small, being around $0.3\%$ in the redshift range $0\leq z\leq 2$. This is of course much smaller than the relative uncertainties of our data points and also than the one of the reconstructed functions (see e.g. our Figs. 2, 8 and 11). Thus, given the sensitivity of the data we are dealing with, the assumption of a flat universe has a derisory impact on our results.

\subsection{Local HST determination of $H_0$}
In this work we also use in some of the analyses the local HST measurement from \cite{RiessH02018}, $H_0^{\rm HST}=(73.45\pm 1.66)$ km/s/Mpc, obtained with the cosmological distance ladder method, using Cepheids and SnIa.

\section{The Gaussian Processes method}\label{sect:GPMethod}

A Gaussian process is a distribution over functions, viz. they generalize the idea of a Gaussian distribution for a finite number of quantities to the continuum. Given a set of Gaussian-distributed data points one can use Gaussian processes to reconstruct the most probable underlying continuous function describing the data, and also obtain the associated confidence bands, without assuming a concrete parametrization of the aforesaid function. Instead, one assumes a concrete kernel, see below and the book \cite{RasmussenWilliams}. This method has been used in several cosmological studies to reconstruct e.g. the DE equation of state parameter \cite{Clarkson2010,Holsclaw2010,Seikel2012} or the expansion history $H(z)$ \cite{Busti2014,VerdeProtopapasJimenez,Li2016,WangMeng,YuRatraWang2017,Melia2018}. In the next subsection we review the main ideas behind the GPs method, and the formulas that are needed to implement it.

\subsection{Reconstruction of functions with Gaussian Processes}

A multivariate normal distribution is defined by a vector of mean values and a covariance matrix. Analogously to it, a Gaussian process is defined by two entities: its mean function, $\mu(z)$, and its two-point covariance function $\mathcal{C}(z,z^\prime)$,
\begin{equation}\label{eq:GP1}
\xi(z)\sim \mathcal{GP} \left(\mu(z),\mathcal{C}(z,z^\prime)\right)\,.
\end{equation}
By definition, a realization $\xi(z)$ of a Gaussian process is a continuous curve, and it is possible to compute the probability of finding a realization inside any region $\xi(z)\pm \Delta\xi(z)$. Let us now focus our attention on the covariance function $\mathcal{C}$ for a while. At those redshifts $z^*$ at which we do {\it not} have any data point, but at which we want to obtain the value of the reconstructed function, we will define $\mathcal{C}(z^*,z^{*\prime})\equiv\mathcal{K}(z^*,z^{*\prime})$. $\mathcal{K}$ is called the kernel function and at this stage it is completely unknown. The only thing that we know about it is that it is a symmetric function that encapsulates the information about the strength of the correlations between the values of the reconstructed function at two different points $z^*$ and $z^{*\prime}$, and also about the amplitude of the deviations from the mean value at a given location. For the points at $\tilde{z}$ at which we {\it do} have data, we know something more. In principle we have access to their observational correlations and uncertainties, and therefore we have their covariance matrix, $D(\tilde{z},\tilde{z}^\prime)$. Of course, the latter can also incorporate the effect of some theoretical uncertainties that might enter the calculation of the final observational output. We define $\mathcal{C}(\tilde{z},\tilde{z}^\prime)\equiv\mathcal{K}(\tilde{z},\tilde{z}^\prime)+D(\tilde{z},\tilde{z}^\prime)$. This is the first step to establish the connection between the data and the underlying function. The correlations between points at a general $z^*$ and a general $\tilde{z}$ are also unknown, so we just define again $\mathcal{C}(z^*,\tilde{z})\equiv\mathcal{K}(z^*,\tilde{z})$. It is obvious that if we want the reconstructed function to vary smoothly, $\mathcal{K}$ cannot be diagonal, i.e. it cannot be just proportional to a Dirac delta, since if it was it would be only able to describe noise at those locations with no data points. There are many options for the kernel function. Three of the most famous ones with only two degrees of freedom are the following: 
\begin{itemize}
\item The Gaussian kernel:
\begin{equation}\label{eq:Gaussiankernel}
\mathcal{K}(z,z^\prime)=\sigma_f^2e^{-\frac{1}{2}\left(\frac{z-z^\prime}{l_f}\right)^2}\,.
\end{equation}
\item The Cauchy kernel:
\begin{equation}\label{eq:Cauchykernel}
\mathcal{K}(z,z^\prime)=\frac{\sigma_f^2l_f}{(z-z^\prime)^2+l_f^2}\,.
\end{equation}
\item The Mat\'ern kernel:
\begin{equation}\label{eq:Maternkernel}
\mathcal{K}(z,z^\prime)=\sigma_f^2\left[1+\frac{\sqrt{3}}{l_f}|z-z^\prime|\right]e^{-\frac{\sqrt{3}}{l_f}|z-z^\prime|}\,.
\end{equation}
\end{itemize}
$\sigma_f$ and $l_f$ are the so-called hyperparameters of the kernel function and have a similar role in the three cases shown above. The first one controls the uncertainties' size and the strength of the correlations, whereas the second somehow limits the scope of these correlations in $z$, i.e. for distances $|z-z^\prime|\gg l_f$ the values of the underlying function at $z$ and $z^\prime$ are very uncorrelated. {\it A priori} one cannot exclude the possibility that the underlying function prefers a kernel which does not only depend on the distance $|z-z^\prime|$, but also on the locations $z$ and $z^\prime$ themselves, in such a way that the symmetry under the interchange $z\leftrightarrow z^\prime$ is kept intact. This is equivalent to say that the hyperparameters could be functions of $z$ and $z^\prime$ as well. We will stick in this work, though, to the more simple kernels \eqref{eq:Gaussiankernel}-\eqref{eq:Maternkernel}. Obviously, the reconstructed function will depend on the values of the hyperparameters and, in principle, also on our choice of the kernel. Once we select the latter, how can we properly select $\sigma_f$ and $l_f$? At this point we must make use of our data. In the GPs philosophy, our data set is conceived as part of a subset of realizations of the Gaussian process. The hyperparameters are chosen so as to maximize the probability of the GP to produce our data set. If we marginalize the GP \eqref{eq:GP1} over the points at $z^*$ we get the following multivariate normal distribution, 
\begin{equation}\label{eq:GP2}
\vec{\xi}\sim \mathcal{N} \left(\{\mu_i(\tilde{z}_i)\},\mathcal{C}\right)\,,
\end{equation}
where $i=1,...,N$, with $N$ being the dimension of the vector of data points $\vec{y}\equiv\{\tilde{z}_i,y_i\}$ at our disposal, and $\mu_i(\tilde{z}_i)$ can be set e.g. to $\vec{0}$ $\forall{i}$, since the result is almost insensitive to this. Thus, the hyperparameters will be obtained upon the minimization of 
\begin{equation}\label{eq:logL}
-2\ln\mathcal{L}(\sigma_f,l_f)=N\ln(2\pi)+\ln|\mathcal{C}(\sigma_f,l_f)|+\vec{y}^T\mathcal{C}^{-1}(\sigma_f,l_f)\vec{y}\,,
\end{equation}
with $\mathcal{L}$ being the marginal likelihood and $|\mathcal{C}|$ the determinant of $\mathcal{C}$. This is the procedure described in \cite{Seikel2012} and used in e.g. \cite{Busti2014,VerdeProtopapasJimenez,YuRatraWang2017} to compute the hyperparameters. Having done this, and using \eqref{eq:GP1}, we can compute the conditional probability of finding a given realization of the Gaussian process in the case in which $\xi(\tilde{z}_i)=y_i(\tilde{z}_i)$. The resulting mean and variance functions extracted from such conditioned GP read, respectively,
\begin{equation}
\bar{\xi}(z^*)=\sum_{i,j=1}^{N}\mathcal{C}^{-1}(\tilde{z}_i,\tilde{z}_j)y(\tilde{z}_j)\mathcal{K}(\tilde{z}_i,z^*)\,,
\end{equation}
\begin{equation}
\sigma^{2}(z^*)=\mathcal{K}(z^*,z^*)-\sum_{i,j=1}^{N}\mathcal{C}^{-1}(\tilde{z}_i,\tilde{z}_j)\mathcal{K}(\tilde{z}_i,z^*)\mathcal{K}(\tilde{z}_j,z^*)\,.
\end{equation}
These are the main outcomes of the GPs method, and will be used repeatedly along this paper. Two comments are in order now: 


\begin{itemize} 
\item The GPs method is in some occasions presented as being model-independent (see e.g. \cite{WangMeng}), and non-parametric in others (see e.g. \cite{Seikel2012}). It is important to qualify these attributes properly. Although it is true that GPs are independent of any cosmological model, they are not model-independent in the technical sense, since one has to choose a specific kernel, which is in charge of controlling the correlations between different points of the reconstructed function. Thus, the shape of the latter can be modified in a greater or lesser extent depending on our particular choice of such kernel, especially at those locations that are very unconstrained by the data. Strictly speaking, the GPs method is also parametric or, to be more precise, hyperparametric. The model-dependent and the parametric nature of the GPs are probably more hidden than in the usual parametric approaches, but these features are still there. So GPs are model-dependent in the sense explained before, and also (hyper)parametric.
\item As recognized in \cite{Seikel2012}, a complete Bayesian analysis would demand to marginalize over the hyperparameters, rather than optimizing Eq. \eqref{eq:logL}. The authors of \cite{Seikel2012} state that in most cases the log marginal likelihood is sharply peaked and that in such cases the optimization is a very good approximation of the marginalization procedure. The authors of \cite{Busti2014,VerdeProtopapasJimenez,YuRatraWang2017} do not marginalize neither, they apply the same methodology as in \cite{Seikel2012}. At this stage we will apply this approximate scheme too, but later on we will explicitly show that the distribution of the hyperparameters is not peaked at all in the problem under study, and that a fully consistent analysis forces us to marginalize them, rather than just assume that their distribution is  Dirac-delta-like. This is equivalent to say that we will take into account the propagation of the hyperparameters' uncertainties, something that has been neglected in the aforementioned previous studies.
\end{itemize}

In the following subsection we apply the GPs method to different data sets in order to reconstruct $H(z)$ and derive the corresponding values of $H_0$.

\subsection{Determination of $H_0$ with GPs, cosmic chronometers, the Pantheon+MCT SnIa compilation, and the local HST measurement}


%
\begin{table}[!t]
\centering
\begin{tabular}{|c|c|c|c|c|}
\hline
             \multicolumn{1}{|c}{Kernel}    & \multicolumn{1}{|c|}{Data set(s)} & \multicolumn{1}{c|}{$H_0$} & \multicolumn{1}{c|}{$d(H_0,H_0^{{\rm HST}})$} & \multicolumn{1}{c|}{$d(H_0,H_0^{{\rm P16}})$} \\ \hline\hline
\multirow{4}{*}{Gaussian \eqref{eq:Gaussiankernel}} & {\small CCH}  & $67.42\pm4.75$& -1.20& +0.10 \\ \cline{2-5} 
                  & {\small CCH+Pantheon+MCT} & $67.06\pm 1.68$ & -2.71 & +0.07\\ \cline{2-5} 
                  & {\small CCH+$H_0^{{\rm HST}}$} & $73.15\pm1.57$ & -0.13 & +3.68\\ \cline{2-5} 
                  & {\small CCH+$H_0^{{\rm HST}}$+Pantheon+MCT} & $71.21\pm1.17$ & -1.10 &+3.23\\ \hline\hline
\multirow{4}{*}{Cauchy \eqref{eq:Cauchykernel}} & {\small CCH} & $69.56\pm5.24$ & -0.71 &+0.50\\ \cline{2-5} 
                  & {\small CCH+Pantheon+MCT} & $67.11\pm1.71$ & -2.66 &+0.10\\ \cline{2-5} 
                  & {\small CCH+$H_0^{{\rm HST}}$} & $73.19\pm1.59$ & -0.11 &+3.67\\ \cline{2-5} 
                  & {\small CCH+$H_0^{{\rm HST}}$+Pantheon+MCT} & $71.24\pm1.18$ & -1.09 &+3.23\\ \hline\hline
\multirow{4}{*}{Mat\'ern \eqref{eq:Maternkernel}} & {\small CCH} & $68.80\pm6.34$ & -0.71 &+0.29\\ \cline{2-5} 
                  & {\small CCH+Pantheon+MCT} & $65.97\pm2.20$ & -2.71 &-0.42\\ \cline{2-5} 
                  & {\small CCH+$H_0^{{\rm HST}}$} & $73.15\pm1.61$ & -0.13 &+3.61\\ \cline{2-5} 
                  & {\small CCH+$H_0^{{\rm HST}}$+Pantheon+MCT} & $71.34\pm1.29$ & -1.00 &+3.08\\ \hline
\end{tabular}
\label{GPresults1}
\caption{Values of $H_0$ and corresponding $1\sigma$ uncertainties in [km/s/Mpc] obtained using the GPs method with different kernels and data sets, by optimizing \protect\eqref{eq:logL} to obtain the values of the hyperparameters. We also show in the last two columns the distances $d(H_0,H_0^{{\rm HST}})$ and $d(H_0,H_0^{{\rm P16}})$, as defined in \protect\eqref{eq:distance}, between each of our determinations and the values reported in \cite{RiessH02018} and \cite{Planck2016}, respectively.}
\end{table}
%


We apply now the methodology explained above with the various GP kernels \eqref{eq:Gaussiankernel}-\eqref{eq:Maternkernel} to reconstruct $H(z)$ from different data sources and infer the value of the Hubble parameter $H_0$ in each of these scenarios. The main outcomes of our analysis are shown in our Table 3. The results obtained by using only cosmic chronometers coincide with those presented in Ref. \cite{YuRatraWang2017}, see their Table 2 (Sample 3\_0). We have used our own numerical programs, whereas the authors of \cite{YuRatraWang2017} use the open-source Python package GaPP \cite{Seikel2012}. No difference can be appreciated between our results and theirs, which is of course good. When only CCH are used in the reconstruction of $H(z)$ it is apparently difficult to obtain a resonant determination of $H_0$ by using the three alternative kernels. Nothing could be further from the truth. Although the central values are not coincident and the differences can be even of two units (compare e.g. the results obtained with the Gaussian and the Cauchy kernels), the three predictions are compatible at $1\sigma$. 
\begin{figure}
\centering
\includegraphics[scale=0.7]{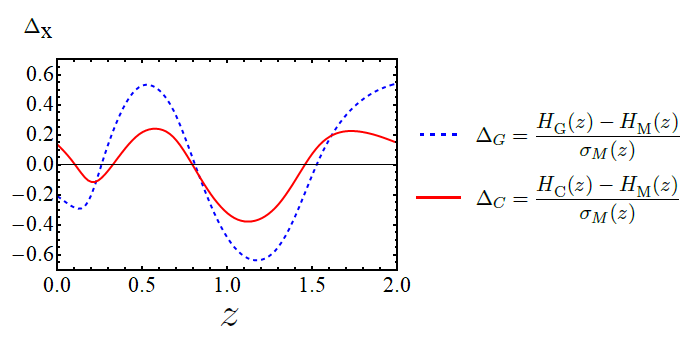}
\caption{Relative differences (as defined in the legend, with X=(G,C)) between the reconstructed functions $H(z)$ obtained with the Gaussian (G) and Mat\'ern (M) kernels (dashed blue curve, $\Delta_G$), and between the ones that are obtained with the Cauchy (C) and the Mat\'ern kernels (solid red curve, $\Delta_C$). The corresponding formulas for the kernels are provided in \protect\eqref{eq:Gaussiankernel}-\protect\eqref{eq:Maternkernel}. Notice that the two curves are below $0.6$ in absolute value, what clearly indicates that the reconstructions with the three kernels are fully consistent (at $<1\sigma_M$). In this case we have only included the CCH data, but the differences are even lower when the Pantheon+MCT data or the $H_0^{{\rm HST}}$ measurement are also considered.}
\label{fig:RelDif}
\end{figure}
This is also manifestly shown in Fig. 1, where we plot the relative differences between the reconstructed functions $H(z)$ obtained with the various kernels. Thus, the results obtained by using only CCH are completely resonant, and seem to prefer the lower range of $H_0$-values advocated in e.g. \cite{TammannReindl2013,Lin2017,Planck2016,PlanckAde}, although they are still compatible at $\sim 1\sigma$ with the measurement from \cite{RiessH02018}, $H_0^\textrm{HST}$, see the last three columns of Table 3. In the last two we collect the values of $d(H_0,H_0^{\rm HST})$ and $d(H_0,H_0^{\rm P16})$, where $d(H_{0,i},H_{0,j})$ is the distance (in $\sigma$ units) between two determinations $H_{0,i}\pm\sigma_i$ and $H_{0,j}\pm\sigma_j$, 
\begin{equation}\label{eq:distance}
d(H_{0,i},H_{0,j})=\frac{H_{0,i}-H_{0,j}}{\sqrt{\sigma_i^2+\sigma_j^2}}\,,
\end{equation}
where $\sigma_{i,j}$ are the corresponding $1\sigma$ uncertainties. These results are telling us that cosmic chronometers cannot pronounce a verdict in favor or against $H_0^{\rm HST}$. They do not have still enough resolution power. This can also be seen by looking at the values of $H_0$ that are obtained by combining CCH and the HST measurement (CCH+$H_0^{\rm HST}$). The CCH data have little impact, and are only able to slightly reduce the original HST determination, both the central value and its uncertainty. Nevertheless the CCH data of Table 1 tend to favor central values lying in the lower range of $H_0$, and only the future will tell us whether the new CCH measurements will persist in favoring this region, by shrinking more and more the error bars while keeping its central value fixed. It is also important to mention that the choice of the SPS model used to extract the CCH data might have a certain impact on our results. This will be analyzed in detail in Sect. 3.4.

\begin{figure}
\centering
\includegraphics[scale=0.8]{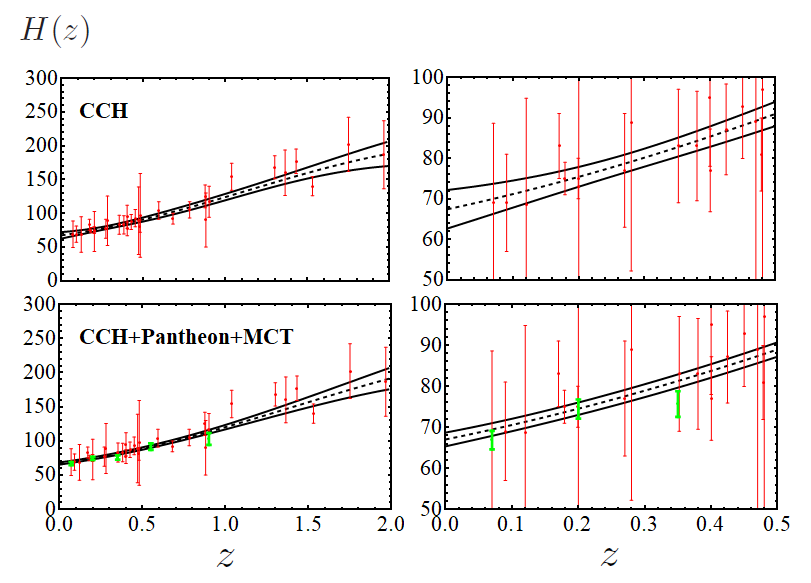}
\caption{Reconstructed $H(z)$ in [km/s/Mpc] with the corresponding $1\sigma$ bands obtained with the GPs method and the Gaussian kernel \protect\eqref{eq:Gaussiankernel} when only CCH are used (the two figures on the top), and when we add the Pantheon+MCT Hubble rates from \cite{Riess2017} (the ones at the bottom). In the two cases we zoom in the redshift range $z\in[0,0.5]$ (see the two plots on the right) in order to better appreciate our determination for $H_0$ and how the uncertainty bands narrow when the SnIa data are included in the analysis. The processed ($E(z_i)\rightarrow H(z_i)$) Pantheon+MCT data obtained in the last iteration of the recursive process are plotted in green, whereas the CCH data (cf. Table 1) are in red. See related comments in the main text.}
\label{fig:H(z)}
\end{figure}


Let us now describe how we have incorporated the Pantheon+MCT compilation data into our analysis. In order to use the Pantheon+MCT data in the reconstruction of the $H(z)$ curve we are forced to translate the Hubble rate data points of Table 2 into Hubble function data. They are related through $H(z)=H_0E(z)$. The linking element is $H_0$, but notice that $H_0$ is precisely what we are looking for. For instance, in the case of the CCH+Pantheon+MCT data set we lack an initial value of the Hubble parameter. In order to solve this problem we have applied an iterative numerical procedure. First of all we make use of the GPs method to derive from (only) the CCH data a value of $H_0$. Then we combine this $H_0$ with the values of the Hubble rate of Table 2 by using a Monte Carlo routine to properly compute the central values and uncertainties of the processed $H(z_i)$. This is how we promote the Hubble rates of the Pantheon+MCT compilation to data points on $H(z_i)$. Then we can apply again the GPs method and derive the ``corrected'' value of $H_0$. We can repeat these steps as many times as we want. It turns out that the resulting value of $H_0$ and its uncertainty converge, so we can decide to stop the iteration process when we attain the desired precision level. In our case we stop the routine when the difference between two successive values of $H_0$ is $\leq 10^{-4}$, which is enough given the uncertainties for $H_0$ shown in Table 3, since they are of order $\mathcal{O}(1)$ in all cases. With this iterative routine we find with the combined CCH+Pantheon+MCT data set central values which are even lower than those found using only CCH (see also Fig. 3). Recall that the Pantheon+MCT data set is obtained by compressing the information of more than $10^3$ SnIa, so it is quite significant that this data set favors the lower range of $H_0$, as the CCH themselves. The values obtained with the three different kernels are again consistent with each other, since they are compatible at $1\sigma$, and their uncertainties are strongly reduced with respect to when only CCH data are considered. To be concrete, the decrease is around a factor $3$, and this makes our determination completely competitive with the HST measurement. The tension between the CCH+Pantheon+MCT value and $H_0^{{\rm HST}}$ is in all three cases of $\sim2.7\sigma$ (cf. Table 3 and Figs. 2-3), so it is quite strong. This result is aligned with the conclusions of Ref. \cite{Lin2017}, in which the authors show that it is highly probable $H_0^{{\rm HST}}$ to be an outlier due to some kind of systematics in the HST measurement. This is also pointed out in \cite{PLB2017}. Other local measurements, though, as the one from Ref. \cite{Bonvin2017}, slightly favor the HST preferred range. Thus the discussion is still open.

\begin{figure}
\centering
\includegraphics[scale=0.7]{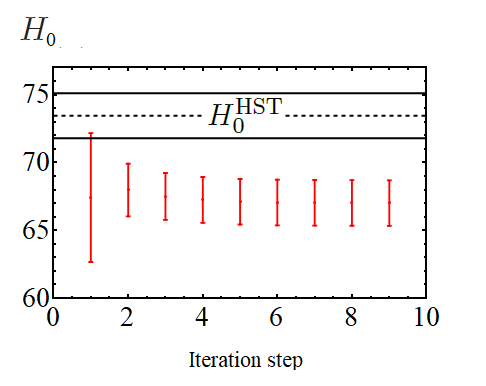}
\caption{Convergence of the CCH+Pantheon+MCT result for the Hubble parameter as a function of the iteration step of the recursive method described in the text (using the Gaussian kernel), together with the value of $H_0^{\rm HST}$ and corresponding $1\sigma$ bands. The values of $H_0$ are given in [km/s/Mpc].}
\label{fig:convergence}
\end{figure}

When the CCH data are combined with $H_0^{{\rm HST}}$, the resulting determination of the Hubble parameter lies very close to $H_0^{{\rm HST}}$ itself, and the uncertainty remains almost unaltered. This is due to the fact that the HST measurement has a low uncertainty and, in addition, it is a direct measurement of $H(z=0)$, so its weight in the GP determination of $H_0$ is greater than the rest of the CCH data points. This also makes the results with the three kernels to be basically the same. Notice that in the CCH+Pantheon+MCT analysis something similar happens when compared with the scenario with only CCH, since by including the Pantheon+MCT data we add points that are very close to $z=0$, and with very low uncertainties, as it is palpable from the two plots at the bottom of Fig. 2. This is basically reflecting what we could have already expected. When we have very precise data the values of the reconstructed function at these points and their vicinity are almost the same, independently of the kernel used in the analysis. This is actually something that must be demanded not only to the GPs but to any reconstruction method, as a matter of consistency. 

Finally, if we consider the full CCH+$H_0^{{\rm HST}}$+Pantheon+MCT data set we obtain again coincident values using the three kernels \eqref{eq:Gaussiankernel}-\eqref{eq:Maternkernel} for the very same reasons explained above. Notice that due to the effect of the Pantheon+MCT data our determination of $H_0$ is now lower than in the CCH+$H_0^{{\rm HST}}$ case, and a small tension appears with $H_0^{{\rm HST}}$ itself, although only at $\sim 1\sigma$. The tension between our determination from CCH+$H_0^{{\rm HST}}$+Pantheon+MCT and Planck's prediction for the $\Lambda$CDM is much larger, reaching in this case the $\sim 3\sigma$ level.

From the study carried out in this subsection we would like to remark two things that we deem important: First, the results obtained with the three different kernels, although are not coincident in some cases, are always consistent with each other. This can be clearly seen by taking a look to Fig. 1 and comparing e.g. the values of the fourth column of Table 3. There we explicitly show the level of discrepancy between our determinations of $H_0$ (obtained with the various data sets) and $H_0^{{\rm HST}}$. The reader can check that for the three kernels used in this work we obtain almost exactly the same predictions once we choose a particular data set; and second, we have shown that the cosmic chronometers used in our analyses and more conspicuously the Pantheon+MCT data set prefer the lower range of values for $H_0$, rather than the region of $H_0^{{\rm HST}}$, and that the tension between the CCH+Pantheon+MCT result and the HST measurement is around the $2.7\sigma$ level.

\subsection{Error propagation of the kernel hyperparameters}
 
The analysis of the previous subsection has been carried out by assuming (as in e.g. \cite{Seikel2012,Busti2014,VerdeProtopapasJimenez,YuRatraWang2017}) that the distributions of the hyperparameters are very peaked, effectively given by Dirac deltas located at the values at which the marginal log likelihood \eqref{eq:logL} is maximum. Now we study which are the changes introduced in our results if we apply the exact procedure by marginalizing the hyperparameters, in the case when only CCH are used. This is the correct way to account for their error propagation. 

First of all, we have drawn the exact distributions of the $\sigma_f$ and $l_f$ hyperparameters from \eqref{eq:logL} by means of a Monte Carlo routine. The result obtained for the Gaussian kernel is presented in Fig. 4. Such distributions are far from peaked, so in principle one could expect some  important corrections to the determinations of $H_0$ shown in Table 3. Similar shapes for the histograms of the hyperparameters are found when the Cauchy or the Mat\'ern kernels are used, so the analysis of this subsection proves to be fully necessary. Our numerical results are presented in Table 4. They have been obtained with a Monte Carlo Metropolis-Hastings algorithm \cite{Metropolis,Hastings} that has allowed us to draw the exact distributions of $H_0$ for the three kernels under use. For each of them we show two different values of $H_0$. The ones on the top correspond to the mean and associated standard deviation extracted from the resulting histograms of $H_0$. The ones on the bottom are the best-fit values and the corresponding uncertainties at $68.3\%$ and $95.5\%$ c.l.. There is an obvious mismatch between the first and the second ones, which is more pronounced for the uncertainties rather than for the central values. The reason is simple. The exact distributions are not exactly Gaussian and this produces these small differences.


\begin{figure}
\centering
\includegraphics[scale=0.5]{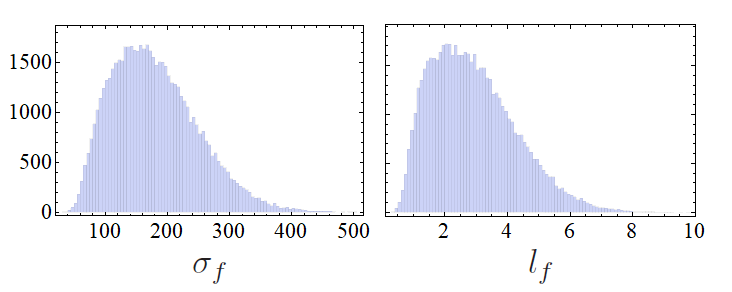}
\caption{Histograms of the hyperparameters $\sigma_f$ (in [km/s/Mpc]) and $l_f$ of the Gaussian kernel \protect\eqref{eq:Gaussiankernel}, obtained by a Monte Carlo sampling of \protect\eqref{eq:logL} and by considering only the CCH data. Both distributions are far from peaked, and this justifies our analysis of Sect. 3.3.}
\label{fig:DistHyper}
\end{figure}


What can we highlight from these corrected results? They are again compatible with each other, so the outputs obtained with the three kernels are consistent. They are also perfectly compatible with the results of Table 3 obtained in the case when only CCH data are used. The corrections introduced by propagating adequately the errors of the hyperparameters are not dramatic at all. It is true that for all three kernels there is a reduction of the central values, which makes our determinations even more resonant with other values that one can find in the literature lying in the lower range of $H_0$, but it is also true that the standard deviations grow as a direct consequence of the propagation of the hyperparameters' errors. This adds an extra uncertainty to the final results, which are still compatible with $H_0^{{\rm HST}}$.

We also want to remark that we cannot add the Pantheon+MCT data points and apply the recursive procedure explained in Sect. 3.2, just because in this case in which we take into consideration the propagation of the hyperparameters' errors we obtain a collection of non-Gaussian-distributed data points for $H(z_i)$ and, therefore, it is not consistent to apply the GPs method, which by definition assume Gaussian-distributed data. This deviation from gaussianity is produced by the one of $H_0$, that we infer in the first step of the iterative process directly from the CCH (see Table 4). The latter breaks the gaussianity of the original Pantheon+MCT Hubble rate data. Despite this, the present analysis has also served to show that the main results of Sect. 3.2 are not strongly altered. Although the shape of the kernel hyperparameters' distributions are not Dirac-delta-like in good approximation, we have explicitly checked that if we marginalize the hyperparameters instead of optimizing the log likelihood \eqref{eq:logL} the conclusions extracted from Sect. 3.2 are kept intact. 

Before ending this section, we would like to study an alternative kernel, with three degrees of freedom, one more than in \eqref{eq:Gaussiankernel}-\eqref{eq:Maternkernel}. It is the so-called rational quadratic kernel,
\begin{equation}\label{eq:3dofkernel}
\mathcal{K}(z,z^\prime)=\sigma_f^2\left(1+\frac{(z-z^\prime)^2}{2\alpha_fl_f^2}\right)^{-\alpha_f}\,.
\end{equation}
\begin{table}[!t]
\centering
\label{GPresults2}
\begin{tabular}{|c|c|}
\hline
                   \multicolumn{1}{|c|}{Kernel}    & \multicolumn{1}{c|}{$H_0$}  \\ \hline
\multirow{2}{*}{Gaussian \eqref{eq:Gaussiankernel}} & $66.31\pm 5.35$ \\ \cline{2-2} 
                  & $66.2^{+4.9+13.4}_{-4.0-10.0}$ \\ \hline
\multirow{2}{*}{Cauchy \eqref{eq:Cauchykernel}} & $68.22\pm5.70$ \\ \cline{2-2} 
                  & $68.0^{+5.0+13.3}_{-4.3-9.6}$ \\ \hline
\multirow{2}{*}{Mat\'ern \eqref{eq:Maternkernel}} & $68.23\pm6.54$ \\ \cline{2-2} 
                  & $68.4^{+5.3+14.8}_{-5.1-11.8}$ \\ \hline
									\multirow{2}{*}{Rational quadratic \eqref{eq:3dofkernel}} & $66.31\pm5.35$ \\ \cline{2-2} 
                  & $66.0^{+4.7+11.4}_{-3.7-9.3}$ \\ \hline
\end{tabular}
\caption{Values of $H_0$ in [km/s/Mpc] obtained using the GPs method and the CCH data set, by propagating properly the hyperparameters' errors instead of minimizing \protect\eqref{eq:logL}. For each kernel we provide two values. The first one corresponds to the mean and standard deviation, whereas the second is the best-fit value and corresponding uncertainties at $68.3\%$ and $95.5\%$ c.l., respectively. The differences between the two values are due to small deviations from gaussianity, see the text in Sect 3.3.}
\end{table}
Apart from $\sigma_f$ and $l_f$ we have now an extra parameter, $\alpha_f$. The latter controls the departure of the kernel \eqref{eq:3dofkernel} from the Gaussian one \eqref{eq:Gaussiankernel}. Note that for very large values of $\alpha_f$ we can Taylor-expand \eqref{eq:3dofkernel} around $z=z^\prime$ by using
\begin{equation}
\lim_{\alpha_f\gg 1}\frac{d^n\mathcal{K}(y)}{dy^n}=(-\alpha_f)^n\sigma_f^2(1+y)^{-\alpha_f}\,,
\end{equation}
where we have defined $y\equiv(z-z^\prime)^2/2\alpha_fl_f^2$. The result reads,
\begin{equation}
\lim_{\alpha_f\gg 1}\mathcal{K}=\sigma_f^2\sum_{n=0}^{\infty}\frac{(-\alpha_f y)^n}{n!}=\sigma_f^2 e^{-\alpha_fy}=\sigma_f^2e^{-\frac{(z-z^\prime)^2}{2l_f^2}}=\eqref{eq:Gaussiankernel}\,.
\end{equation}
Using the GPs method with kernel \eqref{eq:3dofkernel} and the cosmic chronometers data we find $H_0=(64.9\pm4.6)$ km/s/Mpc if we just optimize the log marginal likelihood \eqref{eq:logL}. If, instead, we propagate the errors of the hyperparameters we find the values presented in Table 4. It is remarkable that only in the latter case we recover the results obtained with the Gaussian kernel. This is possible because the CCH data prefer very large values of $\alpha_f$ and, therefore, in this case the limit $\alpha_f\gg 1$ applies, so the rational quadratic kernel \eqref{eq:3dofkernel} just reduces to the Gaussian one \eqref{eq:Gaussiankernel}, and the best-fit values of $\sigma_f$ and $l_f$ coincide with the ones found with \eqref{eq:Gaussiankernel}.

\subsection{Possible systematics affecting the CCH data, and their impact on our results}
The most important source of potential systematic errors affecting the CCH data on $H(z_i)$ comes from the stellar population synthesis model adopted to compute the absolute ages of the passively-evolving galaxies involved in the analyses. For instance, in the works \cite{Moresco2012,Moresco2016} the authors provide 13 $H(z_i)$-values obtained by using two alternative SPS models: the one from Ref. \cite{BC03}, which we call BC03; and the one from Ref. \cite{MaStro}, which we call MaStro. By comparing the BC03 and MaStro values reported in Refs. \cite{Moresco2012,Moresco2016} one can see that 10 out of the 13 measurements are higher when the MaStro SPS model is used, instead of the BC03 one. The differences are not very large, they are lower than $1\sigma$ in all cases except the two at $z=0.7812$ and $z=1.037$, which reach the $1.04\sigma$ and $1.64\sigma$ levels, respectively. Almost all of them are therefore individually compatible, but is important to perform a detailed study of their collective impact on our results. This is the main aim of this subsection.

Let us start describing which SPS models have been used in producing the CCH data points listed in Table 1 and the analyses carried out in the previous subsections. In Refs. \cite{Zhang,Ratsimbazafy2017,Stern} the authors only provide the values of $H(z_i)$ obtained with the BC03 model. This constitutes a $\sim 25\%$ of the whole CCH data set presented in Table 1. In Ref. \cite{Moresco2015} only the combined MaStro/BC03 values are available, whereas in Refs. \cite{Jimenez,Simon} an alternative SPS model is used, different from the MaStro and BC03 ones. These points constitute the $\sim 32\%$ of the CCH data set. In contrast, in Refs. \cite{Moresco2012,Moresco2016} the authors provide both, the BC03 and MaStro values (cf. Tables 4 and 3 of \cite{Moresco2012} and \cite{Moresco2016}, respectively). This group includes the remaining $\sim 43\%$ of the data. We have opted to use the BC03 values of these two references in our main analyses so as to incorporate consistently the data from \cite{Zhang,Ratsimbazafy2017,Stern}, namely to avoid the use of a mixture of MaStro and BC03 values while maximizing the number of data points entering the calculations. Now we want to determine how robust are our results by studying which is their response under different changes of our CCH data set. 

For instance, how do our determinations of $H_0$ change by using the measurements obtained with the MaStro model in Refs. \cite{Moresco2012,Moresco2016} instead of the BC03 ones? Let us stick for simplicity to the results obtained with the Gaussian kernel and without propagating the error of the hyperparameters. If we only use CCH data, we obtain $H_0=(72.11\pm 4.82)$ km/s/Mpc. This has to be compared with the first value of the third column in Table 3. Clearly, there has been an increase of the central value. Now it falls in the HST preferred range, and this shows that there exists a systematic uncertainty comparable to the statistical one. Nevertheless, the difference is not significant from the statistical point of view, since they are both compatible at $<1\sigma$ (at $0.69\sigma$ to be exact). When the Pantheon+MCT Hubble rates are also included, the result is again moved towards lower values, it reads $H_0=(68.93\pm 1.76)$ km/s/Mpc. The tension with $H_0^{\rm HST}$ is now only of roughly $2\sigma$, and is again compatible at $<1\sigma$ with the results obtained using both the CCH measurements derived in the context of the BC03 SPS model and the Pantheon+MCT data (cf. Table 3, second number of the third column). The lowering of $H_0$ is mostly driven by the Hubble rate data points at $z=0.07$ and $z=0.35$ (cf. Table 2), which apart from having a very low relative uncertainty (around $2.5\%$) also acquire low central values, as can be seen in Fig. 1 of Ref. \cite{Riess2017}. If we repeat the test removing these two data points we obtain a much larger value for the Hubble parameter, $H_0=(70.91\pm 2.85)$ km/s/Mpc, as expected. The BC03 and MaStro results clearly favor the lower band of $H_0$-values when the SnIa data are also taken into account. This is also aligned with the results presented in the very recent Ref. \cite{Feeney2018}, in which the authors use BAO data from BOSS together with the Pantheon compilation of SnIa to determine $H_0$. They use the sound horizon at radiation drag inferred by Planck as a prior in their analysis, and find $H_0=(68.57\pm 0.93)$ km/s/Mpc, which is fully resonant with our results, and keeps also the track of other works in the literature, as e.g. \cite{Lin2017,PLB2017,YuRatraWang2017,Addison2017}.

Due to the fact that there is no apparent reason to rely on the MaStro-derived measurements instead of the BC03-derived ones, it is probably better to use the weighted average of the MaStro and BC03 values, and sum the difference in the resulting $H(z)$ in quadrature to the total error, analogously to what is done in Ref. \cite{Moresco2015}. Let us firstly assume that the two SPS models provide completely independent (uncorrelated) results, which of course could be only a first rough approximation. The weighted average of the i$_{th}$ BC03-determination $H_{i,B}\pm\sigma_{i,B}$ and the MaStro one $H_{i,M}\pm\sigma_{i,M}$ is given by
\begin{equation}
\bar{H}_i=\sigma^2_{i,stat}\left(\frac{H_{i,B}}{\sigma_{i,B}^2}+\frac{H_{i,M}}{\sigma_{i,M}^2}\right)\,,
\end{equation}
with
\begin{equation}\label{eq:statErr}
\sigma_{i,stat}=\left(\frac{1}{\sigma_{i,B}^2}+\frac{1}{\sigma_{i,M}^2}\right)^{-1/2}
\end{equation}
being the associated statistical uncertainty. The total uncertainty is obtained upon summing the systematic error, i.e. $\sigma_{i,sys}=|H_{i,B}-H_{i,M}|$, in quadrature to the statistical one, i.e. $\sigma_{i,tot}=\sqrt{\sigma^2_{i,stat}+\sigma^2_{i,sys}}$. The results obtained by using these processed CCH data points with and without the Pantheon+MCT data are $H_0=(67.89\pm 1.79)$ km/s/Mpc and $H_0=(66.20\pm 5.00)$ km/s/Mpc, respectively.

Let us consider now the most unfavorable scenario, in which the BC03 and MaStro $H(z_i)$-values are completely correlated. The generalization of \eqref{eq:statErr} that includes the effect of these correlations is 
\begin{equation}
\tilde{\sigma}^2_{i,stat}=\sigma^2_{i,stat}\left(1+\frac{2\rho_i\sigma_{i,B}\sigma_{i,M}}{\sigma_{i,B}^2+\sigma_{i,M}^2}\right)\,,
\end{equation}
where $\rho_i$ is the correlation coefficient between the i$_{th}$ BC03 and MaStro determinations. Notice that for $\rho_i\to 0$ we retrieve \eqref{eq:statErr}, as desired. The results obtained by assuming $\rho_i=1\,\forall{i}$ with and without the Pantheon+MCT data read now: $H_0=(67.79\pm 1.86)$ km/s/Mpc and $H_0=(66.12\pm 5.16)$ km/s/Mpc, respectively. In this case the statistical uncertainty increases, since now we are losing part of the information carried by the two values due to the existing correlations. Nevertheless, the increase is very small and the results are essentially the same that we obtain by considering $\rho_i=0\,\forall{i}$. Again, no significant deviations from the results presented in Table 3 are found. They prove to be quite robust. When we use only data on $H(z)$ coming from CCH the results are more sensitive to the choice of SPS model, something that was already noted in Ref. \cite{Busti2014}. The determinations of $H_0$ obtained by us in all these checks, though, are fully compatible, and seem to still favor with more power the lower range of $H_0$-values, especially when the SnIa information is also considered. This can be also seen in Fig. 5, where we show the reconstructed curves of $H(z)$ when the BC03, MaStro and the averaged BC03/MaStro data are used in the analysis.

Let us perform yet another check: how do our results change if we {\it only} use data points obtained with the BC03 SPS model, namely if we avoid the use of the CCH data points of Refs. \cite{Jimenez,Simon,Moresco2015}? They read $H_0=(64.43\pm 1.71)$ km/s/Mpc and $H_0=(64.54\pm 4.61)$ km/s/Mpc with and without the Pantheon+MCT data, respectively. Again they are compatible at $\sim 1\sigma$ with the results presented in Table 3.

We have also performed some additional tests. We have studied, for instance, the impact on our results of removing the data points with a relative uncertainty lower than or equal to $10\%$ from the CCH data set of Table 1, i.e. those at $z=0.1791,\,0.1993,\,0.6797,\,1.53$. The results read: $H_0=(69.53\pm 7.96)$ km/s/Mpc and $H_0=(68.33\pm 2.00)$ km/s/Mpc for the CCH and CCH+Pantheon+MCT data sets, respectively. The uncertainties for $H_0$ grow, as expected. We have also determined the Hubble parameter by removing all the data points on $H(z_i)$ from Refs. \cite{Moresco2012,Moresco2016}. The results read now: $H_0=(66.71\pm 6.72)$ km/s/Mpc and $H_0=(67.66\pm 2.24)$ km/s/Mpc for the CCH and CCH+Pantheon+MCT data sets. Again, we can conclude that our determinations remain stable and in any case fully compatible with our previous results. The consistency is kept intact, even when we test the contribution of different redshift regions for the CCH data. The values of $H_0$ obtained with the Pantheon+MCT SnIa data together with the CCH with $z\leq 0.2$ is completely compatible with the one obtained when we consider the CCH data points with $z\leq 0.8$, $z\leq 1.2$ or just take the full CCH data set.

\begin{figure}
\centering
\includegraphics[scale=0.7]{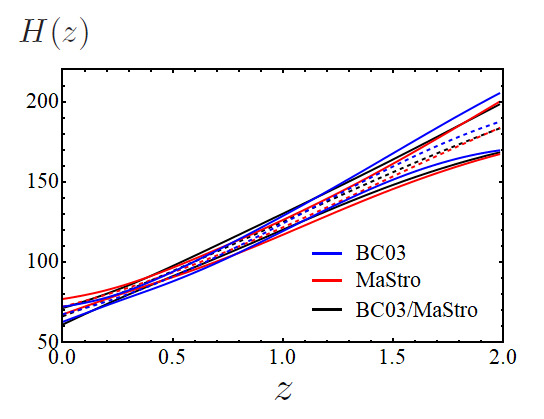}
\caption{Reconstructed curves of $H(z)$ obtained with the GP method (Gaussian kernel) and three different CCH data sets: (i) the one presented in Table 1 (BC03, in blue); (ii) the one that results from exchanging the BC03 data points from Refs. \cite{Moresco2012,Moresco2016} by the MaStro ones (MaStro, in red); and (iii) the one obtained by exchanging the BC03 data with the weighted average of the BC03 and MaStro values from the same references, as described in the text, with $\rho_i=1\,\forall{i}$ (BC03/MaStro, in black).}
\label{fig:SPScomparison}
\end{figure}

The authors of Ref. \cite{VerdeProtopapasJimenez} accounted for the systematics introduced in the analysis by the choice of a particular SPS model removing from their data set (\cite{Simon,Stern,Moresco2012}) those CCH data points at $z>1.2$, which according to \cite{Moresco2012} are the ones with the greatest dependence on the assumed SPS model. They also added a $20\%$ uncertainty to the point at $z = 1.037$. Using GPs they obtained $H_0=(66.2\pm 4.3)$ km/s/Mpc. When the systematics are not taken into account with the complete set of CCH measurements from \cite{Simon,Stern,Moresco2012} the analysis reduces to the one of \cite{Busti2014}, $H_0=(64.9\pm 4.2)$ km/s/Mpc. Both results are compatible and lie also in the lower range.

Another interesting issue we can address is the one concerning the covariance matrix of the CCH data. There is no correlation matrix available in the literature for them, so we have used a diagonal one in our analyses, as described in Sect. 2.1. In order to estimate the potential impact of a hypothetical non-diagonal covariance matrix for the CCH data, we have removed the correlations between the Pantheon+MCT data points and have computed the resulting value of the Hubble function by using the CCH+Pantheon+MCT data set. We deem this is a useful check, since the promoted Hubble function data obtained from the product of the $H_0$ value extracted from CCH and the SnIa Hubble rates have as low relative uncertainties as the most precise CCH data points and, therefore, studying the stability of our results under changes of the covariance matrix in the SnIa sector could give us a good idea of their sensitivity to changes of the covariance matrix in the CCH sector. We have obtained, $H_0=(66.62\pm 2.27)$ km/s/Mpc, which is fully inside the $1\sigma$ region of the result obtained when correlations are duly considered (cf. Table 3). Thus, no dramatic alteration of our results is produced, what again speaks up for their robustness. Despite this, there is no good reason to think that the covariance matrix for the promoted Pantheon+MCT data on $H(z_i)$ is the same or similar to the one for the CCH data. Therefore, this must be conceived only as a first approximate test. 

These tests must be thought of as something useful, but not definitive. We have studied what is the impact of several modifications of our data set on our results and we have seen that, although not exactly the same, they are quite compatible. Of course, the ideal framework would be to have a kind of unique data set with no dispersion coming from the choice of the SPS model. Nevertheless, we have used a quite consistent CCH list (cf. Table 1), in which the $\sim 70\%$ of the points have been obtained with the very same BC03 model, $\sim 5\%$ with a combination of BC03 and MaStro, and $\sim 25\%$ with alternative models. In the future, when the theoretical uncertainties associated to the SPS models are considerably reduced our results will have to be revised. It has also been recently claimed in \cite{LopezCorredoira2017,LopezCorredoira2018} that young stellar components on the quiescent galaxies used to extract the CCH data might have a non-negligible impact on the estimation of $H(z_i)$. When this issue is fully understood also a revision of our results will be needed.


\section{The Weighted Polynomial Regression method}\label{sect:WPR}

Apart from the GPs there exist some alternative methods to reconstruct continuous functions from data. The most common way to do this is to assume a particular theoretical model or parametrization, fit it to the data by using the usual $\chi^2$-minimization to extract the best-fit values of the model parameters and the associated uncertainties, and then propagate them to build the reconstructed function with the corresponding confidence bands. In the case under study, we want to perform such reconstruction without relying on any particular cosmological model nor any particular parametrization. There are in the literature many papers that aim to reconstruct functions from observational data by assuming a concrete parametrization of the quantity of interest, see e.g. \cite{SemizCacimbel2015,MarraSapone2017,vanPutten2018}. This parametrization can have more or less physical motivation, but normally the choice is quite {\it ad hoc} and is just one of the many (if not infinite) possibilities that might exist. Once we choose a particular model or parametrization, the following questions should be answered in a strongly convincing way if we are interested in extracting model-independent information from the reconstructed function: (i) why have we chosen this concrete option, instead of others?; and (ii) how would our results change if an alternative model was selected? It is obvious that, by definition, upon relying on a concrete model in the fitting analysis we cannot derive model-independent conclusions from it. The same is true even if we opt to use a pure phenomenological parametrization, just because all parametrizations implicitly assume some underlying physical behavior or are designed to work better in certain contexts. 

Regression splines can be also useful to reconstruct functions between the range extremes of a given data set without assuming a particular cosmological model, see e.g. \cite{splines}. Notice, though, that in our study we want to obtain $H_0$, which lies outside the cosmic chronometers data range (see Table 1), or precisely in the border if we add the HST data point, $H_0^{{\rm HST}}$, so traditional splines cannot help us to estimate the Hubble parameter in any of these two cases. Smoothed splines are able to alleviate some of the problems, although they are expensive from the computational point of view. Moreover (as in the ordinary splines) we have to choose a particular order for the spline itself, something that is again quite subjective. The weighting method that will be explained later on could be in principle also applied to the smoothed splines approach in order to obtain a definite result from the average of the smoothed splines. In this work, though, we will apply the weighting procedure to a less involved case so as to better illustrate the method without extra complications. 

Principal component analyses are also used to reconstruct functions from data, see e.g. \cite{DEbook} and references therein. Another promising possibility to reconstruct functions in the cosmological context is the so-called cosmographical approach, see e.g. Refs. \cite{CatoenVisser,MortsellClarkson,Aviles2012}. In this framework all the cosmological functions are Taylor-expanded up to a certain order and the coefficients of the resulting polynomials are directly quantities with important physical meaning \footnote{Alternative cosmographical expansions in terms of the Padé or Chebyshev polynomials have also been developed, see e.g. \cite{Gruber2014,Capozziello2017}.}. This approach is quite attractive, since it allows to obtain physical information from data (as e.g. $H_0$ or $q_0$, i.e. the deceleration parameter at the current time) in an almost model-independent way. It only assumes the homogeneity and isotropy of the universe, which is well-supported by observations. From our point of view, though, in this cosmographical method there is still some remaining degree of subjectivity in the choice of the highest order of the Taylor expansion used in the analysis, so although the method does not rely on any particular cosmological model, one must still make a choice in that point. Why is it better to consider all the terms of the Taylor expansion up to order, let us say, $4$ instead of $5$? Why not truncating the series at third order? How do the results change if we move from one option to the other? We will see later on that in the problem under study, i.e. the inference of $H_0$ from the reconstructed curve of $H(z)$, both the value and uncertainty of $H_0$ are quite sensitive to the these issues (most probably because $H_0$ lies at an extreme of our data set), and therefore, we must account for this problem that has been overlooked in past works. This is our main aim in this section.


\begin{figure}[!t]
\centering
\includegraphics[scale=0.6]{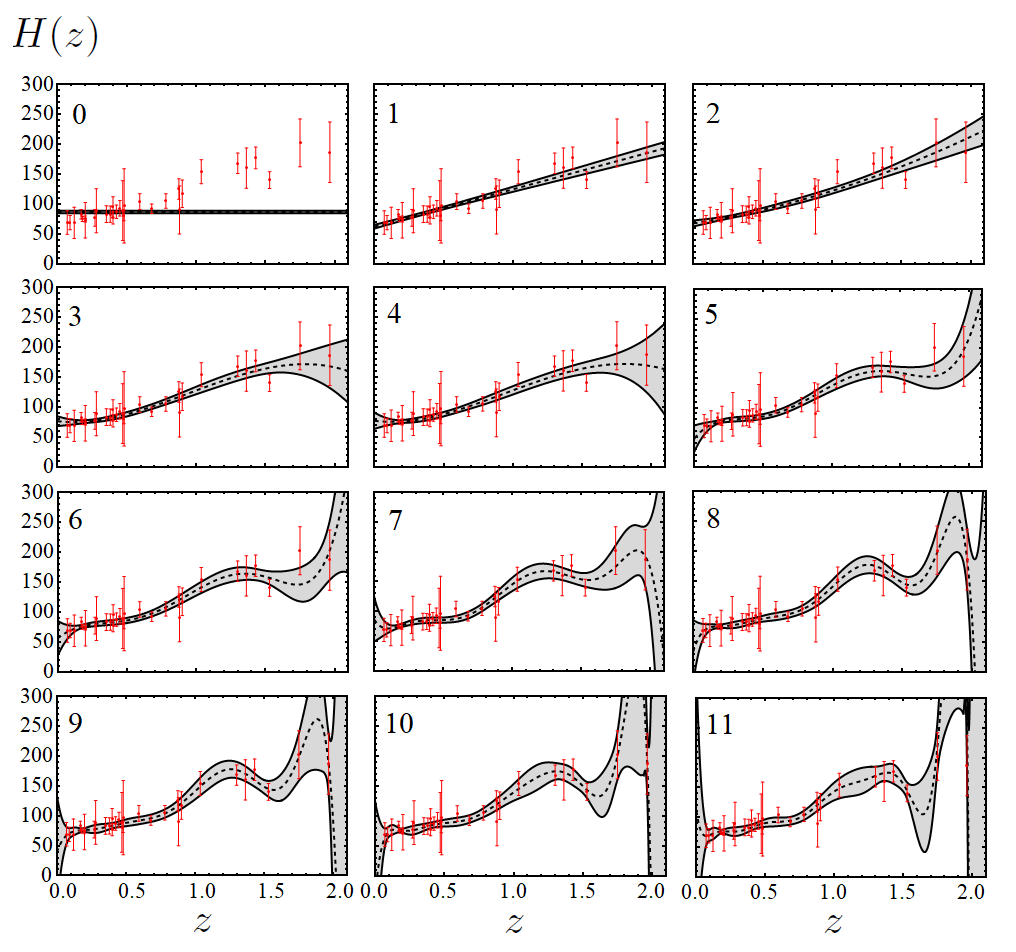}
\caption{Reconstruction of $H(z)$ in [km/s/Mpc] obtained from the CCH data and the cosmographic approach explained in Sect. 4.1 with the twelve polynomials that result from truncating the Taylor expansion at order 1, 2, 3, etc., up to order 12, respectively (the polynomial degree is in each case indicated at the top left corner of the corresponding plot). We have made use of formulas \protect\eqref{eq:covParam}-\protect\eqref{eq:covRecFunc}. Notice that for the polynomials of higher order we need to invert large matrices, and this gives some numerical problems that we have solved by applying the Cholesky decomposition method for matrix inversion.}
\label{fig:CosmographInd}
\end{figure}



\begin{figure}[!t]
\centering
\includegraphics[scale=0.6]{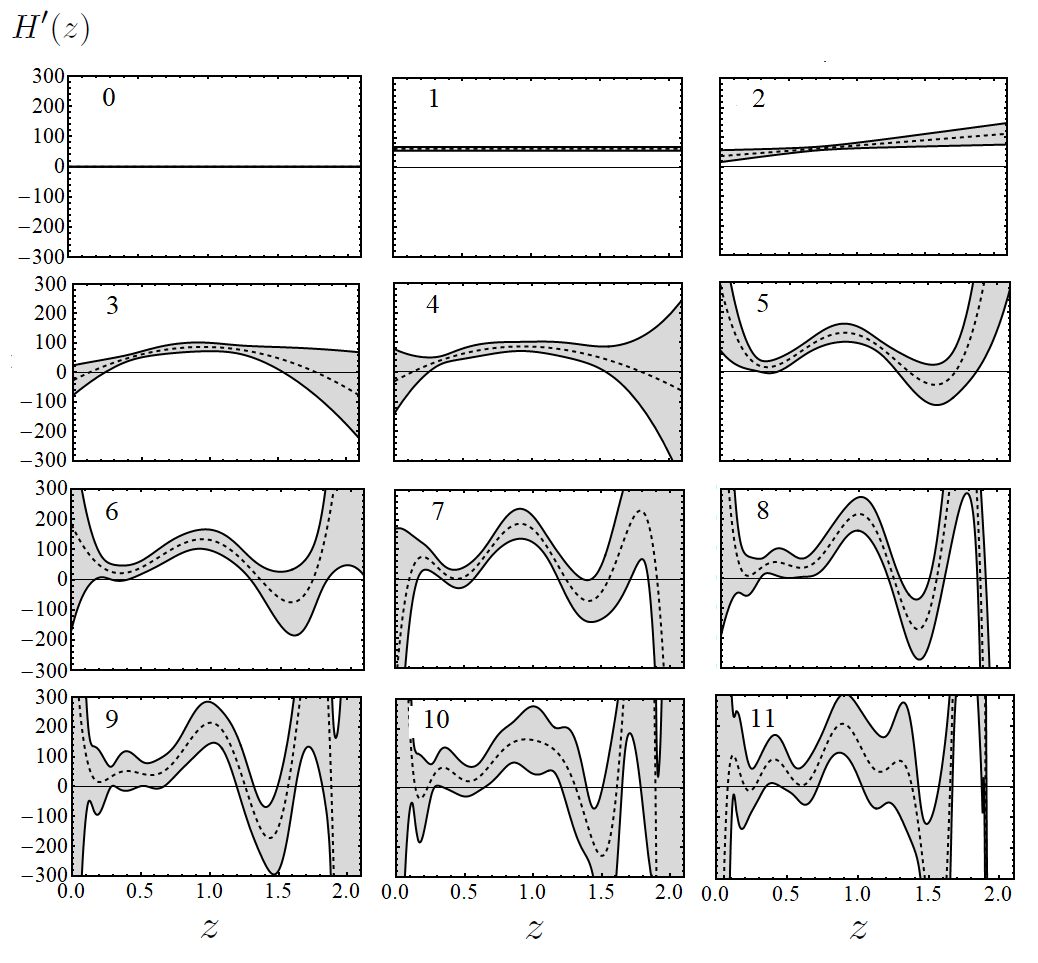}
\caption{As in Fig. 6, but for $H^\prime(z)=dH/dz$. To obtain these plots we have made use of formulas \protect\eqref{eq:covParam}-\protect\eqref{eq:meanParam} and \protect\eqref{eq:meanRecFunc2}-\protect\eqref{eq:covRecFunc2}.}
\label{fig:CosmographIndPrime}
\end{figure}


\subsection{Regression with linear functions in parameters}
In this subsection we quickly review the technique and main expressions needed for fitting data with functions that are linear in the parameters. Imagine we have the following function,
\begin{equation}\label{eq:fitFunc}
f(z)=\sum_{i=0}^{n}a_ib_i(z)\,,
\end{equation}
with $a_i$ being the fitting parameters and $b_i(z)$ some functions of the variable $z$. The latter are the so-called basis functions. If we have a collection $\mathcal{D}=\{(z_\mu,y_\mu),\,\mu=1,...,N,\,N\geq n+1\}$ of Gaussian-distributed data points with covariance matrix $C$, and we want to fit \eqref{eq:fitFunc} to them we have to maximize the likelihood
\begin{equation}\label{eq:likelihood}
\mathcal{L}(\mathcal{D}|\vec{a})=\frac{1}{(2\pi)^{N/2}\sqrt{|C|}}e^{-\frac{1}{2}[y_\mu-f(z_\mu;\vec{a})]C^{-1}_{\mu\beta}[y_\beta-f(z_\beta;\vec{a})]}
\end{equation}
with respect to the elements of the vector of parameters $\vec{a}$. Notice that in the last formula we are using the Einstein summation convention, as we will do in all the forthcoming expressions unless stated otherwise. Also note that we use Greek letters for indexes labeling data points, and Latin ones for those labeling the terms of $f(z)$, as in \eqref{eq:fitFunc}. Due to the linearity of the latter in the parameters it is possible to rewrite the likelihood \eqref{eq:likelihood} as a multivariate Gaussian distribution for the parameters too, i.e.
\begin{equation}\label{eq:likelihood2}
\mathcal{L}(\mathcal{D}|\vec{a})\propto e^{-\frac{1}{2}(a_i-\bar{a}_i)D_{ij}^{-1}(a_j-\bar{a}_j)}\,,
\end{equation}
where
\begin{equation}\label{eq:covParam}
D_{ij}^{-1}=B_\mu^i C_{\mu\beta}^{-1} B_\beta^j\,,
\end{equation}
is the (inverse) covariance matrix of the paramaters, $B_\mu^i\equiv b_i(z_\mu)$, and
\begin{equation}\label{eq:meanParam}
\bar{a}_i=y_\mu C_{\mu\beta}^{-1}B_\beta^jD_{ij}\,.
\end{equation}
According to Bayes' theorem (see e.g. \cite{DEbook}), if the prior distributions for the parameters are constant and flat we can directly identify \eqref{eq:likelihood2} with $\mathcal{L}(\vec{a}|\mathcal{D})$ up to a normalization factor, independent from $\vec{a}$. Thus, in this (and only this) constant flat prior scenario \eqref{eq:meanParam} is the mean (or best-fit) value of the i$_{th}$ parameter. In any case, it is quite straightforward to compute the mean function $\bar{f}_M(z)$ and covariance matrix ${\rm cov}[f_M(z),f_M(z^\prime)]$ once a set of functions $\{b_i\}$ are chosen, i.e. once we select a particular model $M$. It can be done as follows (here we write again the sum symbols explicitly),
%
\begin{table}[!t]
\centering
\label{table5}
\begin{tabular}{|c|c|c|c|c|c|c|}
\hline
 {\small Polyn. degree (d)} & $\chi^2_{{\rm min}}$ & BIC & AIC & $H_0$ & $B_{{\rm d}1}$ (BIC)  & $B_{{\rm d}1}$ (AIC) \\ \hline
 $0$& $122.57$ & $126.00$ & $124.70$ & $86.8\pm2.1$ & $6\times 10^{-23}$ & $3\times 10^{-23}$\\ \hline
 $1$& $16.62$ & $23.48$  & $21.04$ & $62.3\pm 3.1$ & $1$ & $1$\\ \hline
 $2$& $14.74$ & $25.04$ & $21.63$ & $67.8\pm5.1$  & $0.47$ & $0.74$\\  \hline
 $3$& $12.95$ & $26.69$ & $22.49$ & $76.4\pm8.2$ &  $0.20$ & $0.48$\\ \hline
 $4$& $12.95$ & $30.12$ & $25.35$ & $76.9\pm13.5$ & $0.04$ & $0.12$\\ \hline
 $5$& $10.09$ & $30.70$ & $25.59$ & $48.4\pm21.6$ &  $0.03$ & $0.10$ \\ \hline
 $6$& $9.94$ & $33.98$ & $28.81$ & $56.0\pm29.2$ &  $0.005$ & $0.02$\\ \hline
 $7$& $8.15$ & $35.62$ & $30.69$ & $86.3\pm37.0$ &  $0.002$ & $0.008$\\ \hline
 $8$& $6.17$ & $37.08$ & $32.74$ & $33.7\pm52.6$ &  $0.001$ & $0.003$\\ \hline
 $9$& $6.17$ & $40.51$ & $37.17$ & $39.1\pm95.5$ &  $2\times 10^{-4}$ & $3\times 10^{-4}$\\ \hline
 $10$& $5.83$ & $43.61$ & $41.73$ & $-32.7\pm156.5$ & $4\times 10^{-5}$ & $3\times 10^{-5}$\\ \hline
 $11$& $5.57$ & $46.78$ & $46.90$ & $128.1\pm247.8$ & $1\times 10^{-5}$ & $3\times 10^{-6}$\\ \hline
\end{tabular}
\caption{Results obtained from the fitting analysis of the (cosmographic) polynomials up to order 11, by using the CCH data. In the second column we show the minimum value of the $\chi^2$ function, i.e. -2 times the exponent of \protect\eqref{eq:likelihood}. The third and fourth columns contain the values of the Bayesian and Akaike information criteria, as defined in \protect\eqref{eq:criteria}. In the fifth column we show the values of the Hubble parameter in [km/s/Mpc] together with its $1\sigma$ uncertainty. The last two columns contain the values of the Bayes ratio that are obtained by using the BIC and AIC, respectively, as defined in \protect\eqref{eq:Bayesratio}. See comments in the text.}
\end{table}
\begin{equation}\label{eq:meanRecFunc}
\bar{f}_M(z)=\sum_{i=0}^{n}\bar{a}_ib_i(z)\,,
\end{equation}
\begin{equation}\label{eq:covRecFunc}
{\rm cov}[f_M(z),f_M(z^\prime)]=\sum_{i,j=0}^{n} D_{ij}b_i(z)b_j(z^\prime)\,.
\end{equation}
The variance of the reconstructed function is just $\sigma_M^2(z)={\rm cov}[f_M(z),f_M(z)]$. It is also easy to compute the mean function and covariance matrix of the derivatives,
\begin{equation}\label{eq:meanRecFunc2}
<f^{l)}_M(z)>=\sum_{i=0}^{n}\bar{a}_i\frac{d^lb_i(z)}{dz^l}\,,
\end{equation}
\begin{equation}\label{eq:covRecFunc2}
{\rm cov}[f_M^{l)}(z),f_M^{w)}(z^\prime)]=\sum_{i,j=0}^{n} D_{ij}\frac{d^lb_i(z)}{dz^l}\frac{d^wb_j(z)}{dz^w}\,.
\end{equation}
These tools can be used, e.g. to reconstruct $H(z)$ in the cosmographic scenario. We just have to Taylor-expand $H(z)$ around e.g. $z=0$,
\begin{equation}\label{eq:TaylorH}
H(z)=H_0+\frac{dH}{dz}\Bigr\vert_{z=0}z+\frac{1}{2}\frac{d^2H}{dz^2}\Bigr\vert_{z=0}z^2+...
\end{equation}
In this case $b_i(z)=z^i$, and $a_0=H_0$, $a_1=H_0(q_0+1)$, etc. Applying the formulas \eqref{eq:covParam}-\eqref{eq:covRecFunc2} and the CCH data we obtain the results presented in Figs. 6-7 and Table 5 for the twelve polynomials that result from truncating the Taylor expansion \eqref{eq:TaylorH} at order 1, 2, 3, etc., up to order 12, respectively. Let us take a look on Fig. 6 first. It is evident that the shape of the reconstructed functions and $1\sigma$ confidence bands change a lot for the various (cosmographic) polynomials, especially at those regions with only few data points, i.e. far from $z=0$, but also (although at lesser extent) in the low-redshift range. This dispersion is of course also passed on the value of $H_0$ inferred from each of them, see the fifth column of Table 5 too. This clearly is pointing out that choosing a particular polynomial in the cosmographic approach is completely insufficient and unjustified, since it leads us to incompatible determinations of $H_0$. And the trouble is even enhanced for the derivative of $H(z)$, see Fig. 7. Notice that the problem is much more severe than the one affecting the choice of the kernel in the GPs method (see Sects. 3.2 and 3.3). In the next subsection we propose a way to get rid of this drawback.


\subsection{Reconstruction of functions with the WPR method}
To overcome the problem exposed above, it proves useful to regard it from a different angle. We should focus our efforts on trying to solve the following questions: why do we have to rely on the results of a particular polynomial? Does it exist a way of incorporating the information of all the candidate fitting polynomials consistently and, thus, a way of skipping the problem of choosing just one among them? Yes, it does. Let us call $M_0$, $M_1$,..., $M_{N-1}$ the cosmographic polynomials of order $n=0$, $1$,..., $N-1$, respectively. That is, let us conceive each polynomial as a different model, and compute the probability density associated to the fact of having a certain shape for the function $f(z)$ as follows,
\begin{equation}
P[f(z)]=k\cdot[P(f(z)|M_0)P(M_0)+...+P(f(z)|M_{N-1})P(M_{N-1})]\,,
\end{equation}
where $k$ is just a normalization constant that must be fixed by imposing
\begin{equation}
\int[\mathcal{D}f]\,P[f(z)]=1\,.
\end{equation}
Taking into account that 
\begin{equation}
\int[\mathcal{D}f]\,P(f(z)|M_J)=1\quad \forall J\in[0,N-1]
\end{equation}
and  
\begin{equation}\label{eq:normRel}
\sum_{J=0}^{N-1} P(M_J)=1\,,
\end{equation}
we find $k=1$ and therefore:
\begin{equation}
P[f(z)]=\sum_{J=0}^{N-1}P(f(z)|M_J)P(M_J)\,.
\end{equation}
We now denote $M_*$ as the most probable model (later on we will study possible ways to identify it among the whole set $\{M_J\}$) and rewrite the last expression as follows,
\begin{equation}\label{eq:penExp}
P[f(z)]=P(M_*)\sum_{J=0}^{N-1}P(f(z)|M_J)\frac{P(M_J)}{P(M_*)}\,,
\end{equation}
where $\frac{P(M_J)}{P(M_*)}$ can be identified with the Bayes ratio $B_{J*}$ \cite{KassRaftery}. Using \eqref{eq:normRel} one finds
\begin{equation}
P(M_*)=\left(\sum_{J=0}^{N-1}B_{J*}\right)^{-1}\,,
\end{equation}
so \eqref{eq:penExp} can be finally written as
\begin{equation}\label{eq:finExp}
P[f(z)]=\frac{\sum\limits_{J=0}^{N-1}P(f(z)|M_J)B_{J*}}{\sum\limits_{J=0}^{N-1}B_{J*}}\,.
\end{equation}
This is the central expression of the weighted regression method, where the weights are directly given by the Bayes factors. Notice that from \eqref{eq:finExp} we can compute the (weighted) moments and related quantities too. For instance, the weighted mean and variance read,
\begin{equation}\label{eq:MeanTotal}
\bar{f}(z)=\frac{\sum\limits_{J=0}^{N-1}\bar{f}_{J}(z)B_{J*}}{\sum\limits_{J=0}^{N-1}B_{J*}}\,,
\end{equation}
\begin{equation}\label{eq:VarianceTotal}
\sigma^2(z)=\frac{\sum\limits_{J=0}^{N-1}[\sigma_J^2(z)+(\bar{f}_{J}(z))^2]B_{J*}}{\sum\limits_{J=0}^{N-1}B_{J*}}-(\bar{f}(z))^2\,,
\end{equation}
where $\bar{f}_{J}(z)$ and $\sigma^2_J(z)$ can be computed by means of \eqref{eq:meanRecFunc} and \eqref{eq:covRecFunc}, respectively. For the obtention of the Bayes factors we invoke the time-honored Bayesian and Akaike information criteria, BIC and AIC, defined as \cite{Schwarz,Akaike}:
\begin{equation}\label{eq:criteria}
{\rm BIC}_J=\chi^2_{{\rm min},J}+n_J\ln N\qquad {\rm AIC}_J=\chi^2_{{\rm min},J}+\frac{2n_JN}{N-n_J-1}\,,
\end{equation}
with $\chi^2_{{\rm min},J}$ being the minimum of the $\chi^2$ function in the model $M_J$, i.e. -2 times the exponent of \eqref{eq:likelihood} evaluated with the best-fit values $\vec{a}_J$, and $n_J=J+1$ the number of fitting parameters for this model, i.e. the dimension of $\vec{a}_J$. Recall that $N$ is the number of data points entering the analysis. In this way the Bayes factor between the model $M_J$ and the most probable model $M_*$ can be approximated by 
\begin{table}[!t]
\centering
\label{WPRresults1}
\begin{tabular}{|c|c|c|c|c|}
\hline
             \multicolumn{1}{|c}{Weight used}    & \multicolumn{1}{|c|}{Data set(s)} & \multicolumn{1}{c|}{$H_0$} & \multicolumn{1}{c|}{$d(H_0,H_0^{{\rm HST}})$} & \multicolumn{1}{c|}{$d(H_0,H_0^{{\rm P16}})$}  \\ \hline\hline
\multirow{2}{*}{BIC} & CCH  & $65.46\pm 8.36$&  -0.94 &-0.18\\ \cline{2-5} 
                  & CCH+$H_0^{{\rm HST}}$ & $73.17\pm 1.66$ & -0.12 &+3.52\\ \cline{2-5} 
\hline\hline
									\multirow{2}{*}{AIC} & CCH  & $66.79\pm 10.86$&  -0.61 &-0.01 \\ \cline{2-5} 
                  & CCH+$H_0^{{\rm HST}}$ & $73.27\pm 1.66$ & -0.08 &+3.58\\  \cline{2-5}
 \hline
								
\end{tabular}
\caption{Mean values of $H_0$ and corresponding standard deviations in [km/s/Mpc] obtained using the (cosmographic) WPR method with two different kind of weights, constructed with the AIC and BIC \protect\eqref{eq:criteria}, see the main text for details. As in Table 3, we also show in the last two columns the distance from each of our determinations to $H_0^{{\rm HST}}$ and $H_0^{{\rm P16}}$, respectively.}
\end{table}
%
%
\begin{equation}\label{eq:Bayesratio}
B_{J*}=e^{\frac{BIC_*-BIC_J}{2}}\qquad {\rm or}\qquad B_{J*}=e^{\frac{AIC_*-AIC_J}{2}}\,,
\end{equation}
depending on the criterion used, the BIC or AIC, respectively. The most probable model $M_*$ is defined as the one with lowest BIC or AIC, depending again on the chosen criterion. It is crystal-clear from these expressions that the competing models with more parameters used to analyze the same data receive a suitable penalty and, therefore, our weighted method implements in practice Occam's razor principle. Later on we will show that the results obtained by using the BIC and AIC are fully compatible, so our results do not strongly depend on this particular choice. One can even opt to use an alternative criterion, as the deviance information criterion (DIC) \cite{DIC}, defined as ${\rm DIC}=2<\chi^2>-\chi^2_{\rm min}$, where $<\chi^2>$ is the mean value of the $\chi^2$ function, and has to be computed with a Monte Carlo routine. Again, we have explicitly checked that no significant differences from the statistical point of view are obtained in this case, so in this work we will stick to the more traditional Bayesian and Akaike information criteria.  

Let us now look again the numbers in Table 5 and, more concretely, the firsts four and last two columns. It is obvious that a polynomial of order 0, i.e. a constant, is completely unable to properly fit the CCH data, since the $\chi^2_{\rm min}$ and also the BIC and AIC values are more than a hundred units greater than the ones obtained with the other polynomials. This is also evident from Fig. 6 (see the upper leftmost plot) and the last two columns of Table 5, where one can see that this model is a factor $10^{23}$ less probable than the polynomial of order one, and is also very unfavored when compared with the other models regardless of the information criteria used. Polynomials of higher order succeed in lowering the value of $\chi^2_{\rm min}$, just because they have more degrees of freedom and thus are able to fit better the data. Conversely, the BIC and AIC values are maximum when the polynomial is of order 1, so in the problem under study the most probable model is just a straight line. The parabola can also compete quite well, and non-negligible contributions are introduced also by the polynomials of order 3, 4 and 5, see the Bayes ratios for these and the rest of the polynomials in the last two columns of Table 5. In point of fact, it is not necessary to consider the whole set of cosmographic polynomials ($N=31$ in this case) to obtain a good enough reconstructed function $H(z)$ from the CCH data. We only have to take into account the polynomials of order 1-8, since the others only add negligible corrections to the overall result. We continue the discussion on the WPR method in the next subsection. 

\begin{figure}
\centering
\includegraphics[scale=0.65]{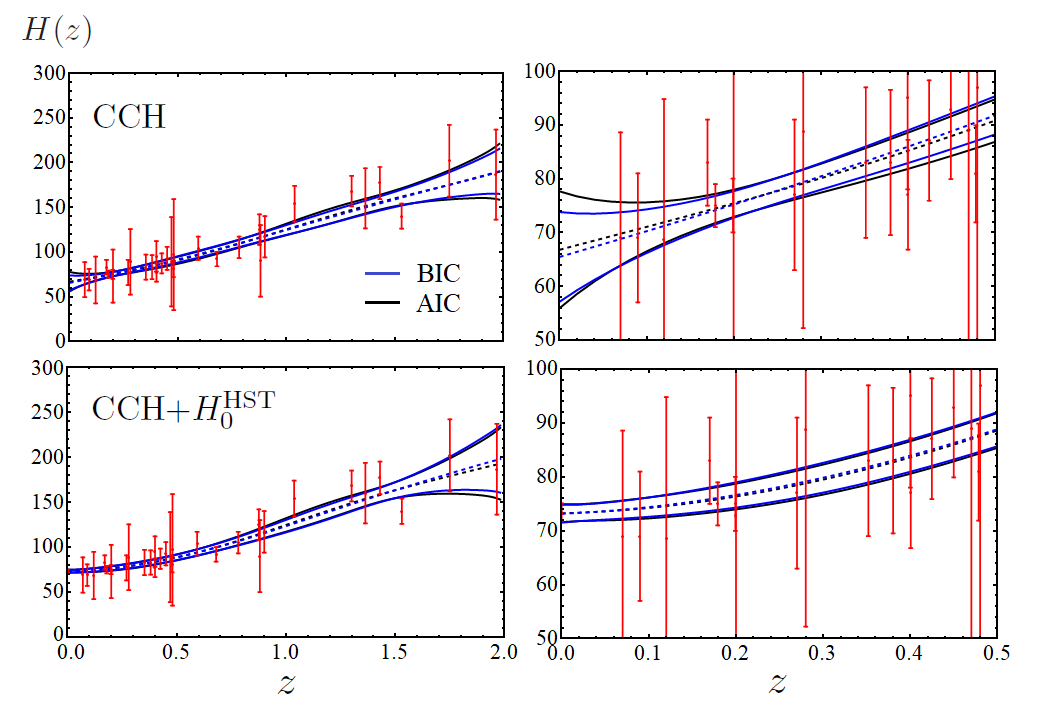}
\caption{Reconstructed $H(z)$ in [km/s/Mpc] with the corresponding $1\sigma$ bands obtained when only CCH are used as input data (the two figures on the top), and when we add the $H_0^{{\rm HST}}$ measurement (the ones at the bottom). In both cases we zoom in the redshift range $z\in[0,0.5]$ (the two plots on the right). See related comments in the main text.}
\label{fig:H2(z)}
\end{figure}
%

\subsection{Constrained WPR, with CCH, SnIa, and $H_0^{\rm HST}$}  
The inferred values of $H_0$ obtained by applying the weighting method using only CCH and also adding $H_0^{{\rm HST}}$ are collected in Table 6. The corresponding reconstructed curves are plotted in Fig. 8. The first thing we would like to remark is that the results obtained with the WPR method in this cosmographic approach are fully compatible with the ones that are obtained with GPs (cf. Table 3). The CCH data favor again the lower band of $H_0$-values, although the results obtained with the WPR method are compatible with the HST measurement at $<1\sigma$ and therefore these differences are not statistically relevant. The uncertainties are now a factor $\sim 1.5-2$ larger than those obtained with GPs, later on we will see why. Again, the effect of the CCH on the results obtained with the combined CCH+$H_0^{{\rm HST}}$ data set is very weak for the very same reasons explained in Sect. 3.2. Moreover notice that the results obtained by applying the Bayesian and Akaike information criteria are fully consistent with each other. This can be checked in Figs. 8 and 9. In the right plot of Fig. 9 we explicitly show that the differences when only the CCH data are considered do not surpass the $0.3\sigma_{{\rm AIC}}$ level, and it is even lower when the HST data point is also included in the analysis.

By direct comparison of the individual plots of Fig. 6 (obtained for each polynomial by using the CCH data) with the final (weighted) curve at the top-left of Fig. 8, it is easy to realize that there exist important differences between them. These differences can be also appraised in terms of $H_0$, by comparing the results of Tables 5 and 6. For instance, if we opted to use a third-order polynomial to reconstruct $H(z)$ (as it is done e.g. in Ref. \cite{MarraSapone2017} with the very same CCH data set employed by us; or in Ref. \cite{vanPutten2018} from a similar compilation of $H(z_i)$ points) we would determine $H_0=(76.4\pm 8.2)$ km/s/Mpc, whereas the BIC-weighted value obtained by us reads $H_0=(65.46\pm 8.36)$ km/s/Mpc, and a similar result is obtained using the Akaike criterion (see Table 6). Although they are both compatible with each other, the first one in centered in the high band of $H_0$-values, whereas the second lies in the lower one. And things worsen even more when polynomials of higher degree are considered. This is why we strongly prefer not to choose a particular polynomial, but use the WPR method to remove the subjectivity that is inherent to the choice of the polynomical degree in the cosmographic approach. There is no {\it adhoc} choice now, since Occam's razor is in charge of selecting the optimal reconstructed function for us.

\begin{figure}
\centering
\includegraphics[scale=0.73]{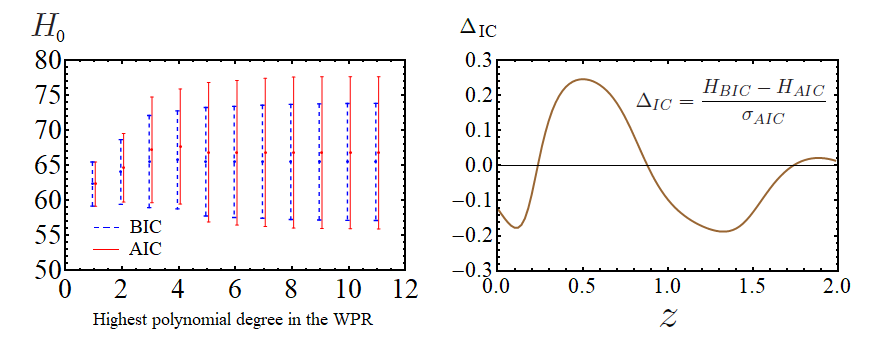}
\caption{{\it Left plot:} Values of $H_0$ in [km/s/Mpc] obtained with the cosmographical WPR approach in the case when only CCH data are considered and using the Bayesian and Akaike information criteria \protect\eqref{eq:criteria} to weight the polynomials. Here we show the convergence of our results according as more terms in the Taylor expansion are introduced in the analysis; {\it Right plot:} Relative difference (as defined in the legend) between the reconstructed functions $H(z)$ when the BIC and AIC are used, in the same scenario as in the left plot. This curve is below $0.3$ in absolute value in all the studied range of $z$'s, so the differences between the two reconstructed functions are really small.}
\label{fig:RelDif}
\end{figure} 

In view of all what we have exposed before, the WPR method seems to work quite well, and this fact seems to be supported by Fig. 8 too, which shows a quite reasonable behavior of $H(z)$. There is still, though, one disturbing issue affecting this cosmographical implementation of the WPR, which we address right now. Figs. 6 and 7 show that many of the polynomials that we are considering lead to an anomalous behavior of the derivative of $H(z)$ (and also of $H(z)$ itself when $z\gtrsim 1$). This is telling us that maybe we are associating a too large probability to unphysical configurations of the polynomials. How can we cure this problem? Instead of considering the expansion \eqref{eq:TaylorH} let us construct an alternative set of polynomials with those fulfilling the following conditions: (i) $H_0\ne 0$; (ii) $H^\prime(z=0)\ne 0$; and (ii) $H^\prime(z)>0\,\forall z>0$, where, recall, the prime denotes a derivative with respect to the redshift $z$, as in Fig. 7. The starting expression from which we will obtain the family of polynomials satisfying these conditions that aim to solve our problem is the following,
\begin{equation}\label{eq:MotherExp}
H^\prime(z)=\alpha\prod_{i=1}^{s}(z-\gamma_i)\,,
\end{equation}
%

\begin{table}[!t]
\centering
\begin{tabular}{|c|c|c|}
\hline
               Polyn. degree (d=s+1)   & $H(z)$ & \# param.\\ \hline
              $1$A   & $a+bz$ & $2$ \\ \hline
              $2$A   & $a+bz+cz^2$ & $3$ \\ \hline
			        $3$A   & $a+bz+cz^2+dz^3$ & $4$ \\ \hline	
						  $3$B   & $a+\alpha b^2z+\alpha bz^2+\frac{\alpha}{3}z^3$ & $3$ \\ \hline 
							$3$C   & $a+bz+cz^3$ & $3$\\ \hline		
							$4$A   & $a+bz+cz^2+dz^3+ez^4$ & $5$ \\ \hline
							$4$B   & $a+4\alpha b^3z+6\alpha b^2z^2+4\alpha bz^3+\alpha z^4$ & $3$ \\ \hline
							$4$C   & $a+\alpha c b^2z+\alpha z^2\left(bc+\frac{b^2}{2}\right)+\frac{\alpha}{3}z^3(c+2b)+\frac{\alpha}{4}z^4$ & $4$ \\ \hline
							$4$D   & $a+\alpha bcz+\frac{\alpha b}{2}z^2+\frac{\alpha c}{3}z^3+\frac{\alpha}{4}z^4$ & $4$ \\ \hline
\end{tabular}
\caption{List of polynomials obtained by requiring the fulfillment of the conditions (i), (ii) and (iii) explained in the text. All the constants appearing in the expressions of the second column are real and positive. The last column indicates the number of free parameters in the corresponding polynomial.}\label{PolyExpr}
\end{table}

\begin{table}[!t]
\centering
\label{table5}
\begin{tabular}{|c|c|c|c|c|c|c|}
\hline
 {\small Polyn. degree (d)} & $\chi^2_{{\rm min}}$ & BIC & AIC & $H_0$ & $B_{{\rm d}1}$ (BIC)  & $B_{{\rm d}1}$ (AIC) \\ \hline
 $1$& $16.62$ & $23.48$  & $21.04$ & $62.3\pm 3.1$ & $1$ & $1$\\ \hline
 $2$& $14.74$ & $25.04$ & $21.63$ & $67.8\pm5.1$  & $0.47$ & $0.74$\\  \hline
 $3$& $14.78$ & $28.51$ & $24.31$ & $68.9\pm4.5$ &  $0.08$ & $0.19$\\ \hline
 $4$& $14.80$ & $31.97$ & $27.20$ & $68.2\pm4.2$ & $0.014$ & $0.05$\\ \hline
 $5$& $14.95$ & $35.56$ & $30.45$ & $68.4\pm4.0$ &  $0.002$ & $0.009$ \\ \hline
 $6$& $15.01$ & $39.05$ & $33.88$ & $68.6\pm4.4$ &  $4\times 10^{-4}$ & $0.002$\\ \hline
 $7$& $15.27$ & $42.74$ & $37.82$ & $68.3\pm4.2$ &  $6\times 10^{-5}$ & $2\times 10^{-4}$\\ \hline
 $8$& $15.72$ & $46.63$ & $42.29$ & $68.8\pm4.4$ &  $9\times 10^{-6}$ & $2\times 10^{-5}$\\ \hline
 $9$& $16.38$ & $50.72$ & $47.38$ & $68.9\pm3.9$ &  $1\times 10^{-6}$ & $2\times 10^{-6}$\\ \hline
\end{tabular}
\caption{As in Table 5, but for the constrained cosmographic functions up to order 9, by using again only the CCH data. As explained below in the text, we force the coefficients of these polynomials to be strictly positive by using the prior $a_i>0$ for all the parameters.}
\end{table}
\noindent with $\alpha$ being a positive real in order to make possible the fulfilling of our condition (iii) at very large values of $z$, since $H^\prime(z)\sim\alpha z^{s}$ in this limit. After integrating \eqref{eq:MotherExp} one obtains $H(z)$, a polynomial of degree $s+1$, with constant term $H(z=0)\ne 0$ to ensure our condition (i). The analysis of the zeros of \eqref{eq:MotherExp} is pivotal. If we want to satisfy (iii) we must demand the coefficients $\gamma_i$ to be negative or complex, and in the latter case it is compulsory that the terms in \eqref{eq:MotherExp} that contain complex numbers compensate each other producing only polynomials with real coefficients. This is only possible if the complex terms appear in combinations like $(z-j\gamma_i)^{n}(z+j\gamma_i)^{n}=(z^2+\gamma_i^2)^n$, where $n$ is a natural number and $j$ is in this case the imaginary unit. Upon the application of these conditions (ii) is automatically satisfied, and we obtain the polynomials presented in Table 7. We only show there the polynomials up to order 4, just to illustrate our discussion. For instance, in the case in which $s=3$ and thus $H(z)$ is a polynomial of order 4, we obtain 4 polynomials with different functional behavior (cases A, B, etc.). The polynomial 4A is obtained when the three roots of \eqref{eq:MotherExp} are real and distinct; 4B when they are real and the same; 4C when only 2 are equal, but all are again real; and 4D when two are complex (one conjugate from the other) and the other one is real. A similar reasoning leads to the expressions for the other polynomials listed in Table 7. Notice, though, that the polynomial 3B is a particular case of 3A, and that the polynomials 4B, 4C and 4D are particular cases of 4A. On the other hand, 3C does not acquire a cosmographical form, since it lacks the quadratic term. Thus, if we restrict ourselves to the cosmographic approach, then we end up only with the A-type polynomials, which have the same structure as the ones studied before, but with the particularity that the coefficients are now strictly positive.

We can therefore repeat our previous analysis by just applying a prior $\Pi(a_i)    n\propto\Theta(a_i)$ (with $\Theta(a_i)$ being the Heaviside step function) for the coefficients in order to force them to be positive. The individual mean values and variances for the various constrained cosmographic polynomials cannot be computed with formulas \eqref{eq:meanParam} and \eqref{eq:covParam}, respectively, because now we are not using constant flat priors as before, but step ones, which obviously alter the results. The calculation has to be carried out numerically by means of a Monte Carlo routine. We present in Table 8 the analogous of Table 5 for the constrained cosmographic WPR analysis. Thanks to the changes introduced in it, the values of $H_0$ derived from each polynomial are more reasonable and stable than those obtained with the previous ones, i.e. the statistical variance associated to the set of values of the Hubble parameter tabulated in Table 8 is much lower than the one in the previous study (cf. Table 5). It is also quite illustrative to compare the reconstructed functions of Fig. 6 with those of Fig. 10. The final results obtained with these constrained functions and the WPR approach are shown in Table 9, and graphically in Fig. 11. The values extracted from the CCH+$H_0^{{\rm HST}}$ data are almost identical to the ones derived with the cosmographical polynomials without using any prior for the parameters (see Table 6 for comparison), but when only the CCH data are used one important difference appears with respect to the previous results. Although the central values are very similar and they are actually fully compatible, the current uncertainties are strongly reduced by a factor $\sim 2$. This is due to the fact that by imposing the aforesaid priors we are introducing extra information in the analysis, which helps us to constrain better the parameters. The uncertainties are now similar to those obtained with the GPs method, and a $1.5-2\sigma$ tension with $H_0^{{\rm HST}}$ arises. In the rightmost plots of Fig. 11 we can also appreciate the reconstructed shape of the derivative $H^\prime(z)$, which of course remains positive. Unfortunately, the precision at which $H^\prime(z)$ is obtained does not allow us to provide an accurate determination of the deceleration-acceleration transition redshift, as it is also concluded in Ref. \cite{YuRatraWang2017} in the context of the GPs method.


%
\begin{table}[!t]
\centering
\label{WPRresults2}
\begin{tabular}{|c|c|c|c|c|}
\hline
             \multicolumn{1}{|c}{Weight used}    & \multicolumn{1}{|c|}{Data set(s)} & \multicolumn{1}{c|}{$H_0$} & \multicolumn{1}{c|}{$d(H_0,H_0^{{\rm HST}})$} & \multicolumn{1}{c|}{$d(H_0,H_0^{{\rm P16}})$} \\ \hline\hline
\multirow{2}{*}{BIC} & CCH  & $64.44\pm4.72$& -1.80 & -0.52 \\ \cline{2-5} 
& CCH+Pantheon+MCT* & $68.90\pm 1.96$ & -1.77 & +0.96\\ \cline{2-5}
                  & CCH+$H_0^{{\rm HST}}$ & $73.06\pm1.61$ & -0.17&+3.55\\ \cline{2-5} 
\hline\hline
									\multirow{2}{*}{AIC} & CCH  & $65.27\pm4.98$&  -1.56&-0.33 \\ \cline{2-5} 
									& CCH+Pantheon+MCT* & $68.89\pm 1.96$ & -1.78 & +0.95\\ \cline{2-5}
                  & CCH+$H_0^{{\rm HST}}$ & $73.06\pm1.61$ & -0.17 &+3.55\\  \cline{2-5}
 \hline
								
\end{tabular}
\caption{The same as in Table 6, but now using the constrained cosmographic WPR method described in Sect. 4.3, and also including the results obtained with the data set CCH+Pantheon+MCT*, which also incorporates the data point $E^{-1}(z=1.5)$.}
\end{table}
%


\begin{figure}
\centering
\includegraphics[scale=0.7]{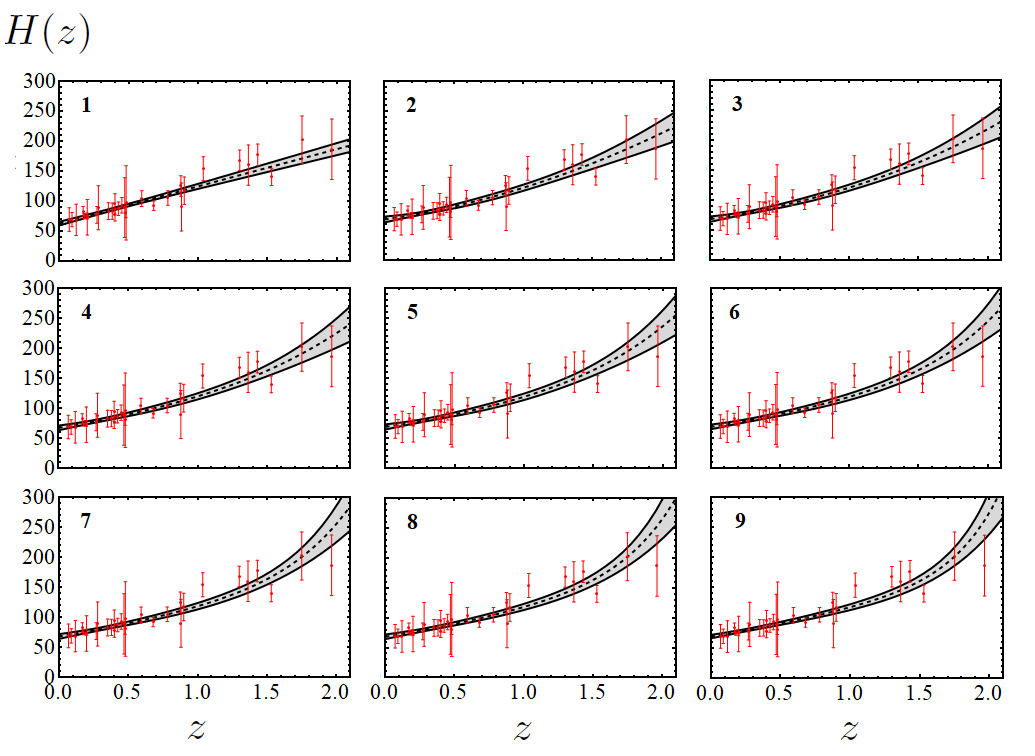}
\caption{Reconstructed curves of $H(z)$ in [km/s/Mpc] obtained with the constrained cosmographic polynomials as described at the end in Sect. 4.3, up to order 9. In contrast to the previous analysis (cf. Fig. 6), the resulting curves are monotonic, as expected.}
\label{fig:CosmographInd2}
\end{figure}

\begin{figure}
\centering
\includegraphics[scale=0.6]{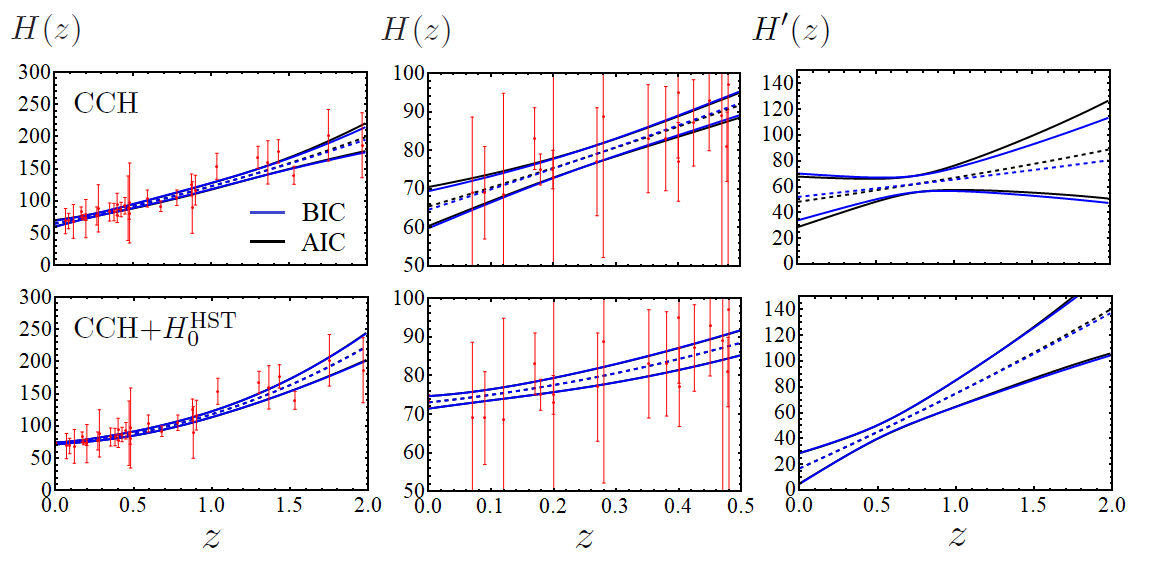}
\caption{Reconstructed $H(z)$ in [km/s/Mpc] with the corresponding $1\sigma$ bands obtained with the constrained WPR method when only CCH are used as input data (the three figures on the top), and when we add the $H_0^{{\rm HST}}$ measurement (the ones at the bottom), using the BIC and AIC to penalize the use of extra degrees of freedom. In both cases we zoom in the redshift range $z\in[0,0.5]$ (the two plots at the center). We also show on the right the plots of the corresponding reconstructed $H^\prime(z)$.}
\label{fig:H3(z)}
\end{figure} 

In Table 9 we also present the result that is obtained with the constrained WPR method and the CCH+Pantheon+MCT* data set, $H_0=(68.90\pm 1.96)$ km/s/Mpc. It is compatible with the values derived from GPs (cf. Table 3). For instance, the distance \eqref{eq:distance} between the latter and the GP determination obtained with the Gaussian kernel, $H_0=(67.06\pm 1.68)$ km/s/Mpc, is only of $0.7\sigma$. Two important questions should be addressed, though: (i) What is the origin of the difference between the uncertainties of $H_0$ obtained with the WPR and the GPs?; and (ii) How can we explain the observed shift in the corresponding central values? In order to answer the first one it should be good to know which of the two uncertainties is most expected. Taking into account that the effective number of free parameters entering the reconstruction procedure in the WPR is roughly 2 (since the polynomial of order one carries the major part of the weight in the weighted sum) we expect to find in the WPR method a similar error for $H_0$ as the one encountered in the model-dependent analysis of the $\Lambda$CDM, in which one also plays with two free parameters. The fitting analysis of the concordance model has been carried out in Appendix A (see in particular Table 11). We find $H_0=(69.07\pm 1.86)$ km/s/Mpc, so the uncertainty (and central value) in the WPR method resonates quite well with the one of the $\Lambda$CDM. In the GP formalism one can compute the effective number of free parameters as indicated in Chapter 2 of Ref. \cite{RasmussenWilliams}. It is basically given by the trace of the product of $\mathcal{K}\mathcal{C}^{-1}$ (see Sect. 3.1), which gives a number slightly lower than 3 and therefore larger than the number of hyperparameters entering the kernel. Thus, we would expect to find in this case a similar uncertainty as in the simple XCDM DE parametrization \cite{XCDM}, in which again an uncertainty of $H_0$ around 1.9 km/s/Mpc is obtained (cf. Table 11). In contrast, we obtain an uncertainty a $\sim 10\%$ lower using GPs. The reason might be the following. In order to incorporate the Pantheon+MCT data to our analysis we have applied an iterative routine, as described in Sect. 3.2. We basically multiply the value of $H_0$ inferred in the preceding GP iterative step with the Pantheon+MCT Hubble rates without considering any correlation between them and $H_0$, and promote the Hubble rates to values on $H(z_i)$. Then we apply the GPs again in order to obtain the updated value of $H_0$ without considering the correlations between the CCH and SnIa data. Then we go to the next iterative step. This procedure is of course almost exact in the first step, but not in the subsequent ones because in these subsequent steps the correlations between the CCH and SnIa data become important. So, if we were more conservative and stopped the iterative procedure at step one in order to avoid the underestimation of the existing correlations, what result would we obtain? Using the Gaussian kernel we would obtain $H_0=(67.99\pm 1.94)$ km/s/Mpc, which is even more compatible with the WPR determination and seems to have a more reasonable uncertainty. Using the latter value and the WPR one we can also derive a GPs+WPR result, by using \eqref{eq:MeanTotal} and \eqref{eq:VarianceTotal}, and weighting both methods equally. Doing this we finally obtain $H_0=(68.45\pm 2.00)$ km/s/Mpc, which is still almost $2\sigma$ away from the local HST determination. In Table 10 we present a summary of all these results.


\section{Discussion and conclusions}\label{sect:DiscussionConclusions}
In this paper we have reconstructed the Hubble function using Gaussian processes (GPs) and the novel weighted polynomial regression (WPR) method from a cosmographical perspective. We have studied the impact on these reconstructions of the most updated cosmic chronometers (CCH) and supernovae of Type Ia (SnIa) data from the Pantheon+MCT compilation and have determined the value of the Hubble parameter that is inferred from them. By using the Gaussian kernel we have found $H_0=(67.06\pm 1.68)$ km/s/Mpc. Thus, the combined CCH+Pantheon+MCT data set favors the lower range of $H_0$-values, as the CMB measured by Planck \cite{Planck2016,PlanckAde} and other many cosmological data sets \cite{Addison2017,Lin2017,Feeney2018}. In order to incorporate the Hubble rate points extracted from the Pantheon+MCT SnIa into the GPs calculation and use them together with the CCH data on $H(z_i)$ we have made use of the iterative procedure described in Sect. 3.2. Stopping the iterative routine at step one yields a more conservative result, $H_0=(67.99\pm 1.94)$ km/s/Mpc. We have checked that the Pantheon+MCT Hubble rate data points, which compress the information of more than one thousand SnIa, are very constraining and are able to reduce in a factor $\sim 3$ the uncertainty of $H_0$ with respect to the case when only CCH data are considered. This is remarkable, since the resulting value of $H_0$ obtained from the combined CCH+Pantheon+MCT data set has now only a $2.5-3\%$ uncertainty (depending on the kernel used, cf. Table 3), which is directly comparable to the one from the local (distance ladder) measurement carried out by the Hubble Space Telescope (HST) \cite{RiessH02018}, $H_0^{{\rm HST}}$. The tension with the latter reaches the $2-3\sigma$ level. This fact seems to support the idea of regarding $H_0^{{\rm HST}}$ as an outlier, see the exhaustive analysis of Ref. \cite{Lin2017}, and also \cite{Addison2017,PLB2017,Feeney2018}.

\begin{table}[!t]
\centering
\begin{tabular}{|c|c|c|}
\hline
               Method   & Data set & $H_0$ \\ \hline
\multirow{3}{*}{GPs (Gaussian kernel \eqref{eq:Gaussiankernel})} & CCH & $67.42\pm 4.75$ \\ \cline{2-3} 
                  & CCH+Pantheon+MCT (full convergence) & $67.06\pm 1.68$ \\ \cline{2-3} 
									& CCH+Pantheon+MCT (1 iteration)  & $67.99\pm 1.94$ \\ \hline
\multirow{2}{*}{Constrained WPR (BIC)} & CCH & $64.44\pm 4.72$ \\ \cline{2-3} 
                  & CCH+Pantheon+MCT* & $68.90\pm 1.96$ \\ \hline
\multirow{2}{*}{GPs+WPR} & CCH & $65.93\pm 4.96$ \\ \cline{2-3} 
                  & CCH+Pantheon+MCT & $68.45\pm 2.00$  \\ \hline
\end{tabular}
\label{summary}
\caption{Summary table with the values of $H_0$ obtained with the two reconstruction methods under study in this work and using CCH with and without the SnIa information. We also provide the combined GPs+WPR values, see the text for further comments. The values of $H_0$ are given in [km/s/Mpc].}
\end{table}

We have also studied the impact of the error propagation of the hyperparameters' errors in the GPs analysis. We have checked that although the distribution of the hyperparameters is far from peaked (cf. Fig. 4), when this feature is taken into account the output of the analysis is completely compatible with the one that is obtained when this error propagation is omitted. For instance, the corrected value of $H_0$ obtained with CCH and the Gaussian kernel reads $(66.2^{+4.9}_{-4.0})$ km/s/Mpc at $68.3\%$ c.l. (cf. Table 4 for the results obtained with other kernels and data sets). This analysis was clearly needed in order to rely on our previous (uncorrected) results, and also on other studies, as e.g. those presented in \cite{Seikel2012,Busti2014,YuRatraWang2017,VerdeProtopapasJimenez}. 

In Sect. 3.4 we have analyzed the impact that the choice of the stellar population synthesis (SPS) model has on our results. The CCH data seem to be sensitive to it, although the observed differences between the values inferred from the two SPS models under scrutiny (BC03 and MaStro) are in almost all cases lower than $1\sigma$ (cf. Refs. \cite{Moresco2012,Moresco2016}). This study was already carried out in \cite{Busti2014}, but we have deemed necessary to repeat it in the light of the measurements from Ref. \cite{Moresco2016}, which where not available four years ago, when Ref. \cite{Busti2014} appeared. Our conclusion is that the values obtained for $H_0$ are quite robust, especially when the Pantheon+MCT information are also considered together with the CCH. Our results tend to favor the lower range of $H_0$-values.

We have also explicitly checked that the results that are obtained with the GPs method with the Gaussian \eqref{eq:Gaussiankernel}, Cauchy \eqref{eq:Cauchykernel} and Mat\'ern \eqref{eq:Maternkernel} kernels are fully consistent (cf. Tables 3 and 4, and Fig. 1), and also that when we have more data around a point the prediction of these three kernels seem to converge in its vicinity, which is of course reassuring. We have also studied the behavior of an alternative kernel with three hyperparameters, the rational quadratic one \eqref{eq:3dofkernel} (cf. Table 4). We have seen that in the problem under study it basically reduces to the Gaussian kernel, with only two degrees of freedom.

We have also presented a new alternative method to reconstruct functions by properly weighting the fitting results obtained with a finite set of cosmographic polynomials. This allows us not to rely on a particular cosmographic expansion for the fitting function or a concrete kernel in the GP formalism. In this sense, the method removes some part of the existing degree of subjectivity affecting in greater or lesser extent the original cosmographical approach, and also the GPs. We have applied the method from two different cosmographical approaches and have obtained consistent results which resonate quite well with those obtained with the GPs. We have also provided a unified GPs+WPR value for the Hubble parameter, $H_0=(68.45\pm 2.00)$ km/s/Mpc. This result is very similar to the one obtained from BAO and the Pantheon SnIa in \cite{Feeney2018}, $H_0 = (68.57\pm 0.93)$ km/s/Mpc. The uncertainty is a factor two lower in this case with respect to our  combined result, basically due to the constraining power of the BAO data. Nevertheless it is important to remark that they use the sound horizon at radiation drag inferred by Planck (assuming the $\Lambda$CDM) as a prior in their analysis. The effect of such model-dependent prior, though, should be small, as it is argued by the authors. Our combined result is also compatible with other ``model-independent'' less-constraining estimates of $H_0$ available in the literature, as e.g. \cite{ChenRatra2011,Busti2014,VerdeProtopapasJimenez,YuRatraWang2017,WangMeng}.

The Hubble parameter $H_0$ was the first ever measured cosmological parameter, but almost ninety years after Hubble's first measurement we still lack a univocal value of it. Although there are other local values supporting the higher range, as e.g. those from \cite{Bonvin2017,WangMeng2017}, the tension between the HST determination \cite{RiessH02016,RiessH02017,RiessH02018} and other local measurements \cite{TammannReindl2013,OnSandage}, and also other sources of cosmological data (see e.g. \cite{Lin2017}), is there and has not been cured yet. We have shown that not only the CCH, but also the SnIa data from the Pantheon+MCT compilation prefer lower values of $H_0$, in contrast to the HST determination. There is a $2-3\sigma$ tension between the CCH+Pantheon+MCT value and $H_0^{{\rm HST}}$. Of course, more data and more studies on the possible systematics affecting the various data sources will be needed to resolve the origin of this tension and to determine whether $H_0^{{\rm HST}}$ is or not an outlier, and whether new physics is needed or not to explain the true value of the Hubble parameter. In the meanwhile we can confirm that $H_0^{{\rm HST}}$ seems to be in conflict with a large variety of alternative data, including the CCH data listed in Table 1 and the SnIa information from Ref. \cite{Riess2017}.


\acknowledgments AGV is grateful to the Institute of Theoretical Physics of the Ruprecht-Karls University of Heidelberg for the financial support and hospitality during the firsts stages of the elaboration of this paper. The work of LA is supported by the DFG through TRR33 ``The Dark Universe''. The authors want to express also their gratitude to Prof. Joan Sol\`a for reading this manuscript and for illuminating discussions on the $H_0$-tension; also to Prof. Adam G. Riess, Prof. Licia Verde, and Dr. Daniel L. Shafer for fruitful feedback on the potential impact on our results of the stellar population synthesis models that are used to obtain the CCH data.


\appendix
\section{Model-dependent results for the $\Lambda$CDM and XCDM}
We dedicate this appendix to present some model-dependent fitting results obtained by using the CCH data of Table 1 and the Pantheon+MCT Hubble rates of Ref. \cite{Riess2017} for the concordance $\Lambda$CDM model (see e.g. \cite{DEbook} and references therein) and the simple XCDM DE parametrization \cite{XCDM} (also known as $\omega$CDM). In both cases we consider a flat universe, i.e. we take $\Omega_k^{(0)}=0$. Although it is of course possible to build larger and more constraining data sets by including e.g. CMB or large-scale structure information, we opt to stick to the same data used in the main body of this paper, since this will serve us to compare the results obtained for these well-known models with those presented throughout this work, which are almost free of model-dependent assumptions.  

The XCDM parametrization probably constitutes one of the simplest extensions of the standard $\Lambda$CDM model. Here the rigid cosmological term, $\Lambda$, is replaced by a DE entity $X$ with equation of state (EoS) $p_X=\omega \rho_X$, with $\omega$ being the constant EoS parameter. $X$ is self-conserved and therefore if $\omega\ne -1$, i.e. if we depart from the concordance scenario, its energy density evolves with the expansion as follows,
\begin{equation}\label{eq:rhoX}
\rho_X(z)=\rho_X^{(0)}\,(1+z)^{3(1+w)}\,.
\end{equation}
where $\rho_X^{(0)}=\rho_X(0)$. Thus, the Hubble function in terms of the redshift is given by
\begin{equation}\label{eq:HXCDM}
H(z)=H_0\sqrt{\Omega^{(0)}_m\,(1+z)^{3}+(1-\Omega^{(0)}_m)\,(1+z)^{3(1+w)}}\,,
\end{equation}
with $\Omega^{(0)}_m=\rho_m^{(0)}/(\rho_m^{(0)}+\rho_X^{(0)})$. Notice that in \eqref{eq:HXCDM} we have neglected the contribution of the relativistic species, since $\rho_r(z)/\rho_m(z)\lesssim 10^{-3}$ in the redshift range covered by the data, i.e. $z_i\in(0,2)$, and their addition would only introduce relative corrections $\lesssim 0.1\%$ in the computation of $H(z)$, and this is much lower than the relative uncertainties of the data points (cf. Tables 1-2). Notice also that by setting $\omega=-1$ we automatically retrieve the $\Lambda$CDM results, i.e. $\rho_\Lambda(z)=\rho_X^{(0)}$ and the corresponding Hubble function from \eqref{eq:rhoX} and \eqref{eq:HXCDM}, respectively.

\begin{table}[!t]
\centering
\begin{tabular}{|c|c|c|c|c|}
\hline
Model & Data set(s) & $H_0$ [km/s/Mpc] & $\Omega^{(0)}_m$ & $\omega$ \\ \hline
 \multirow{3}{*}{$\Lambda$CDM}  & CCH & $68.14\pm 3.07$ & $0.320\pm 0.059$ & -1 \\ \cline{2-5}  
 & CCH+Pantheon+MCT & $69.30\pm 1.88$ & $0.294\pm 0.021$ & -1 \\ \cline{2-5}
& CCH+Pantheon+MCT* & $69.07\pm 1.86$ & $0.299\pm 0.021$ & -1 \\ \hline
\multirow{3}{*}{XCDM} & CCH & $71.74\pm 7.76$ & $0.321\pm 0.057$ & $-1.33\pm 0.62$ \\ \cline{2-5}
& CCH+Pantheon+MCT & $69.16\pm 1.89$ & $0.319\pm 0.047$ & $-1.08\pm 0.15$ \\ \cline{2-5}
& CCH+Pantheon+MCT* & $68.94\pm 1.87$ & $0.328\pm 0.045$ & $-1.10\pm 0.15$ \\ \hline

\end{tabular}
\caption{Fitting results obtained for the $\Lambda$CDM and XCDM by using different data sets. The CCH+Pantheon+MCT* one does not only include the data points listed in Tables 1 and 2, but also the Pantheon+MCT data point at $z=1.5$ \cite{Riess2017}, which has not been used in the GP-reconstruction analyses of $H(z)$ performed in the main sections of this work for the reasons exposed in Sect. 2.2 and the caption of Table 2. The comparison of the results obtained with CCH+Pantheon+MCT and CCH+Pantheon+MCT* shows the low impact of this data point on the overall results.}
\label{ModelDependent}
\end{table}

The fitting results for the two models under study are shown in Table 11. These are some aspects we would like to remark: 

\begin{itemize}
\item The addition of the Pantheon+MCT SnIa information to the CCH data allows us to reduce the uncertainty of $H_0$ a factor $\sim 2$ in the context of the $\Lambda$CDM, and a factor $\sim 4$ in the context of the XCDM. This confirms the constraining power of this data set (recall that by using the GPs method in Sect. 3 and the WPR in Sect. 4 we have been able to reduce the uncertainty of $H_0$ by a factor $\sim 3$ in these more model-independent approaches).

\item The preference of the Pantheon+MCT data for the lower range of $H_0$-values is manifest also in the framework of both, the $\Lambda$CDM and XCDM. When only the CCH data are considered this preference is diluted due, in part, to the increase of the error bars, which makes the determination compatible or only mildly in tension with $H_0^{\rm HST}$.

\item When the CCH+Pantheon+MCT* data set is used the uncertainties of the best-fit values of $H_0$ are only $\sim 1\%$ lower than when we use the CCH+Pantheon+MCT data set. In the former we are also adding the Hubble rate at $z=1.5$ of Ref. \cite{Riess2017}. In the GP analyses we have avoided the use of this data point, just because it gives rise to a non-gaussianly distributed value of $H(z=1.5)$, but in this model-dependent study we can use it without any problem, as when we use the WPR method. By comparing the results obtained with the CCH+Pantheon+MCT and CCH+Pantheon+MCT* it is clear that the impact of this data point on our results is not very important, namely it only produces a decrease in the central value of $H_0$ around the $0.3\%$. The reasons are explained in the caption of Table 2.

\item Regarding the values of $\omega$ in the XCDM, they tend to favor the phantom region, i.e. $\omega<-1$, but the sign is not well fixed, since the uncertainties are too large to be conclusive. They basically allow $\omega$ to be in the region $-1.25\leq\omega\leq -0.95$ at $1\sigma$ c.l. when the CCH+Pantheon+MCT* data set is used. 

\item The derived constraint on $\Omega^{(0)}_m$ is worse in the XCDM than in the $\Lambda$CDM, just because in this model we have an extra free parameter ($\omega$) which absorbs partially the constraining power of the other two when we also include the SnIa information. This absorption is much more enhanced for $\Omega^{(0)}_m$ rather than for $H_0$. The uncertainty in the former is roughly a factor $2$ larger in the XCDM than in the $\Lambda$CDM when the Pantheon+MCT or Pantheon+MCT* data sets are considered together with the CCH.
\end{itemize}

The results shown in this appendix can be compared with those obtained from other analyses of the $\Lambda$CDM and XCDM that are available in the literature, which use similar data sets, as the ones presented in Refs. \cite{Lukovic2016,ChenKumarRatra2017}. They obtain comparable results, fully compatible with ours. See also Ref. \cite{CardassianFit}, where the authors use $H(z_i)$-values from CCH and BAO together with the JLA compilation of SnIa (which is a subset of the Pantheon one, see Sect. 2.2) to constrain the so-called Cardassian models. Curiously, they also obtain values lying in the lower range of $H_0$ when a flat prior for this parameter is used. We want to remark, though, that the analysis carried out in this appendix, and those presented in these references are strongly model-dependent, and so are the derived determinations of $H_0$. In the main body of our work we have estimated $H_0$ using GPs and the WPR method, which do {\it not} rely on any cosmological model. Thus, the spirit of the main work has been quite different from the one of this appendix. It must only be thought of as a support material, presented as a baseline to which compare the model-independent results of the main body of the paper, and other results that one can find in the literature.


\end{document}